\documentclass[sigconf]{acmart}

\usepackage{multirow}
\usepackage{url}
\usepackage{acmart-taps}
\usepackage{pdflscape}
\usepackage{colortbl}
\usepackage{nicefrac}
\usepackage[utf8]{inputenc}

\definecolor{blueF}{rgb}{0.0, 0.0, 0.0}
\definecolor{mypink}{RGB}{242, 216, 228}
\definecolor{myblue}{RGB}{216, 235, 242}
\definecolor{mygreen}{RGB}{229, 242, 216}

\hypersetup{colorlinks=true, citecolor=Navy, linkcolor=blue, urlcolor=blue}

\AtBeginDocument{%
  }
    
\copyrightyear{2025}
\acmYear{2025}
\setcopyright{cc}
\setcctype{by}
\acmConference[CHI '25]{CHI Conference on Human Factors in Computing Systems}{April 26-May 1, 2025}{Yokohama, Japan}
\acmBooktitle{CHI Conference on Human Factors in Computing Systems (CHI '25), April 26-May 1, 2025, Yokohama, Japan}\acmDOI{10.1145/3706598.3714098}
\acmISBN{979-8-4007-1394-1/25/04}

\begin{document}


\title{AI Mismatches: Identifying Potential Algorithmic Harms Before AI Development}


\author{Devansh Saxena$^*$$\dagger$}
\affiliation{%
  \institution{University of Wisconsin-Madison}
  \city{Madison, WI}
  \country{USA}}
\email{devansh.saxena@wisc.edu}

\author{Ji-Youn Jung$^*$}
\email{jiyounj@andrew.cmu.edu}
\affiliation{%
  \institution{Carnegie Mellon University}
  \city{Pittsburgh}
  \state{PA}
  \country{USA}
}

\author{Jodi Forlizzi}
\affiliation{%
  \institution{Carnegie Mellon University}
  \city{Pittsburgh, PA}
  \country{USA}}
\email{forlizzi@cs.cmu.edu}

\author{Kenneth Holstein}
\affiliation{%
  \institution{Carnegie Mellon University}
  \city{Pittsburgh, PA}
  \country{USA}}
\email{kjholste@andrew.cmu.edu}

\author{John Zimmerman}\thanks{\textbf{$^*$ Co-first authors contributed equally to this research. \newline $\dagger$ Author conducted this research as a Presidential Postdoctoral Fellow at Carnegie Mellon University.}}
\affiliation{%
  \institution{Carnegie Mellon University}
  \city{Pittsburgh, PA}
  \country{USA}}
\email{johnz@andrew.cmu.edu}

\renewcommand{\shortauthors}{Saxena and Jung et al.}

\begin{abstract}
AI systems are often introduced with high expectations, yet many fail to deliver, resulting in unintended harm and missed opportunities for benefit. We frequently observe significant "AI Mismatches", where the system’s actual performance falls short of what is needed to ensure safety and co-create value. These mismatches are particularly difficult to address once development is underway, highlighting the need for early-stage intervention. Navigating complex, multi-dimensional risk factors that contribute to AI Mismatches is a persistent challenge. To address it, we propose an AI Mismatch approach to anticipate and mitigate risks early on, focusing on the gap between realistic model performance and required task performance. Through an analysis of 774 AI cases, we extracted a set of critical factors, which informed the development of seven matrices that map the relationships between these factors and highlight high-risk areas. Through case studies, we demonstrate how our approach can help reduce risks in AI development.
\end{abstract}




\maketitle


\section{Introduction}

There has been growing interest and an increasing amount of hype about how AI might revolutionize different industries. The recent emergence of large, pre-trained models (e.g. Large Language Models, Generative AI, Foundational Models) has only accelerated this interest, investment, and hype. 
AI now impacts numerous sectors, including manufacturing, pharmaceuticals, agriculture, and more. Currently, 90\% of CEOs plan to prioritize AI \cite{Kell2024-xw}. AI can offer value through improved insights, decision-making, and resource optimization.

What sometimes gets lost in the hype is the tremendously high failure rate for AI initiatives within organizations (85\%) \cite{Akula2021-vj}, and one of the main causes of failure is the unintended harms these systems can create \cite{weiner2022ai}. For instance, after OpenAI's ChatGPT launch in November 2022, Meta's Galactica was quickly shut down due to poor quality \cite{Heaven2022-fo}. Similarly, a demo error in Google's Bard caused a \$100 billion drop in market value in one day \cite{Olson2023-ht}. The rapid pace of innovation and hype often lead to the release of AI systems that unintentionally cause significant harm.

Over the last decade, the Responsible AI (RAI) and HCI communities developed various 
approaches to reduce AI’s unintended harms. These approaches include fairness toolkits, auditing methods, algorithmic ``de-biasing'' techniques, and documentation protocols~\cite{deng2023understanding,dutta2020there,gebru2021datasheets,holstein2019improving,kallus2018residual,metaxa2021auditing,mitchell2019model,wong2023seeing}. Interestingly, most of these efforts focus on fixing harmful AI systems after they have been deployed or after models have been developed. Recently, researchers have recognized that some AI concepts\footnote{Here, an AI concept refers to an early-stage idea for a specific application of AI, such as an AI system that approves or denies citizen's applications for public benefits by inferring the likelihood of fraudulent behavior based on data like credit reports and social media activity.} are fundamentally flawed --- doomed before the first line of code is written. 
These failures often result from \textbf{AI Mismatches}: mismatches between the \textit{required model performance} needed to deliver the intended value and minimize harm and the \textit{model performance that can be realistically achieved} given factors such as data quality and the complexity of the inference~\cite{coston2023validity, Raji2022-ls, Wang2024-cj}. Moreover, ethical problems can be exceedingly challenging to address once a model is created. Post-hoc fixes can sometimes even exacerbate biases and introduce new ethical harms~\cite{Cooper2021-ft, dutta2020there, kallus2018residual}. 
Thus, there is a growing consensus that RAI concerns are often best addressed at the earliest stages of AI innovation; during the ideation and problem formulation stages of AI development~\cite{coston2023validity,kawakami2024situate,passi2019problem, Raji2022-ls,selbst2019fairness, Wang2024-cj,yildirim2023creating}.  

Recent RAI research maps out various types of AI-related harms and has developed taxonomies to help practitioners anticipate and minimize these harms 
(e.g., \cite{barocas2017problem, wang2022measuring, Shelby2023-ff}). Similarly, recent work in HCI indicates that many successful AI systems create value for users by focusing on simple tasks where it is realistic to expect excellent model performance, or by addressing situations where even moderate model performance creates value for users \cite{yildirim2023creating}. 
For example, moderate performance in automatic speech recognition is sufficient for generating voicemail transcripts, as users can generally understand the text despite some errors. By contrast, moderate performance is inadequate for transcribing court cases, where precise wording is crucial. 

In making the case for thinking about unintended harm at the earliest innovation stages, researchers note that algorithmic harms often arise from the interplay of multiple factors~\cite{Wang2024-cj, marconi2024challenges}. This complicates efforts to address harms using one-dimensional frameworks. Navigating and predicting the factors that increase the risk of AI-related harm, especially during the early stages of development, remains a major challenge for innovation teams.

To address this, we developed an approach for thinking through \textit{AI Mismatches} in early-stage AI concepts to identify the potential for ethical harm. Following this approach, we envision how AI innovators are sensitized to notice mismatches between (i) the model performance we can realistically expect, and (ii) the performance required to provide actual benefit and avoid unintended harm \cite{Raji2022-ls, Wang2024-cj}. We adopt a human-centric definition of model performance \cite{parasuraman2000model, yang2019unremarkable, lai2021towards}. Rather than understanding performance solely in terms of standard ML metrics such as predictive accuracy, we define model performance holistically in terms of the \textit{model's ability to perform a task that fulfills a human need}.

Our approach leverages Yildirim et al.'s Task-Expertise x AI-Performance matrix as a building block \cite{yildirim2023creating}. It helps to explore several perspectives around unintended harms for a given concept. Our approach surfaces mismatches, increasing the ability to foresee potential harms and consider ways to reduce these risks before system building begins. To develop our approach, we reviewed 774 case studies of real-world algorithmic harm, identifying critical underlying risk factors and their interplay with model performance. We then created a set of seven matrices to expose the interrelationships and identify high- and low-risk areas. Finally, we conducted a preliminary assessment of our process by analyzing six case studies involving automated decision-making, generative AI, and machine learning applications. This provides a preview of how our approach might help surface key risk factors during the concept selection phase of an innovation process.

This paper makes four contributions. First, it offers an initial step towards understanding the multi-dimensional factors that influence AI Mismatch. Our matrices serve as an example to help uncover discrepancies that indicate an AI concept might be infeasible to implement in a way that both creates value and minimizes harm. Second, our approach provides explanatory power to articulate inherent risk factors for an AI concept. We believe this can support teams in steering their concepts toward ‘safer’ zones, where a more balanced approach can be taken between what a concept aims to achieve and what is realistically possible. Third, this paper contributes a set of comparative case studies where we used our approach to uncover key AI Mismatch factors that influence the gap between AI capabilities and task requirements. Finally, this paper identifies areas for future study related to AI Mismatch, particularly around design for moderate model performance.

\section{Related Work}

\subsection{AI as a Design Material}\label{subsec:AI_designmaterial}
Over the past decade, some HCI research has explored the concept of AI as a Design Material \cite{dove2017ux, holmquist2017intelligence, yildirim2022experienced, benjamin2021machine}. Researchers reasoned that integrating design thinking into AI innovation would open up the space for innovation by bringing more creative thinking to envisioning what might be built \cite{dove2017ux}. In reaction to the high failure rate of AI initiatives, these researchers suggested the problem might be what innovation teams chose to build in the first place. They noted that in many cases, there seemed to be a lack of ideation and time devoted to exploring many different concepts early on (e.g., sketching, brainstorming, and ideation). They noted that most AI-focused resources and guidebooks only address challenges that happen during the prototyping phase, well after problem formulation and project selection.

In response to this challenge, researchers developed various resources, tools, and processes that enable designers to engage with AI’s potential, envision how AI capabilities can address specific problems, and create value for users and service providers (e.g., \cite{fiebrink2010wekinator, yildirim2023creating, jansen2023mix}). Despite these efforts, findings show that design teams often struggle to identify situations where AI capabilities can realistically create value. Prior work examined why AI is "uniquely difficult to design for" \cite{yang2020re}, with AI’s inherent uncertainty identified as a core challenge. Additionally, some scholars have noted that AI is often metaphorically described as `magical' or `enchanting' because it seems to perform tasks previously thought impossible for computers \cite{lupetti2024making}. The positioning of AI as magic may partly explain why design and HCI seem to overestimate what AI can do and underestimate its costs and harms.

To address the issue of designers overestimating AI's capabilities, Yildirim et al. \cite{yildirim2023creating} introduced the Task-Expertise x AI-Performance matrix. This matrix helps innovation teams more easily recognize low-risk and high-value concepts when dealing with many possible things to build. The matrix consists of three rows representing levels of task expertise. 
For example, a step counter (the task of noticing and counting steps) requires minimal task expertise, while recognizing cancer from a pathology image requires significant expertise. The matrix also has three columns indicating the minimum level of AI model performance needed for a user to experience the system as valuable. For example, automatic speech recognition of a voicemail only needs moderate performance to be useful, while automatic transcription of a court case would need excellent model performance to be useful. 

While this framework holds promise for helping teams generate more feasible AI concepts during ideation, it does not consider the ethical risks embedded in the concepts. Many ethical failures in AI stem from unrealistic problem formulations, so balancing performance requirements with feasibility presents an important area for further research. Our paper builds on this work by identifying high-risk areas within the AI design space, making these risks more visible during the design process.

\subsection{Responsible AI and FATE Research}\label{subsec:RAI_FATE}

Fairness, Accountability, Transparency, and Ethics (FATE) in sociotechnical systems has seen significant growth over the past decade with the emergence of new research communities and conferences (e.g., FAccT and AIES). Given that these research communities were initially dominated by machine learning researchers and legal scholars, much of the early FATE research focused on formalizing specific mathematical definitions of `fairness', along with creating algorithmic techniques that attempt to align existing datasets or AI models to comply with those definitions \cite{selbst2019fairness}. For instance, algorithmic fairness mitigation or `de-biasing' techniques typically rely upon existing datasets both to correct for unfairness and bias and to assess whether such corrections have been `successful'~\cite{Cooper2021-ft, dutta2020there, kallus2018residual}. They assume flawed AI systems should be fixed while never asking what developers should choose not to build.

However, recent findings reveal a paradox: when datasets are severely biased, de-biasing methods can sometimes inadvertently amplify the very biases they aim to correct \cite{Cooper2021-ft, dutta2020there, kallus2018residual}. Consequently, researchers have drawn attention to the fact that FATE concerns can be inherent to a specific problem formulation, requiring a fundamental AI system redesign rather than post hoc adjustments to models or datasets \cite{boyarskaya2020overcoming, holstein2019improving, passi2019problem, Raji2022-ls}. Moreover, some applications—particularly in high-stakes public sector contexts like child welfare—pose unavoidable risks. For example, early AI innovations aimed at predicting child maltreatment risk faced criticism due to the high costs of errors, prompting a shift toward lower-risk, preventive systems \cite{Eubanks2018-dk, Kawakami2022-ez, Stapleton2022-eb, saxena2024algorithmic}. 

In sum, most RAI tools and processes for AI practitioners have largely mirrored the focus on "making the thing right" rather than "making the right thing" \cite{buxton2010sketching}. These efforts focus on refining existing systems or documenting their limitations, as opposed to ideating and choosing better things to make \cite{holstein2019improving, wong2023seeing}. Moreover, research investigating industry product teams' current practices and challenges around AI fairness, found that teams were most interested in finding ways to avoid FATE challenges in the first place \cite{deng2023understanding, holstein2019improving, lee2024don}. To address this, several recent calls to action urge FATE researchers to turn their attention toward the earliest stages of the AI innovation process (e.g., \cite{boyarskaya2020overcoming, holstein2019improving, passi2019problem, Raji2022-ls, wang2023designing}). Our study contributes to this dialogue by showing how a design perspective can help teams preemptively unpack risk factors, identify high-risk regions, and refocus AI practitioners on designing the right thing in the first place.

An emerging dialogue within the FATE community on "AI functionality" highlights how harms often arise when systems underperform, when they make unexpected errors. For instance, Raji et al. \cite{Raji2022-ls} 
highlight that current AI FATE discussions frequently assume systems function as intended, focusing on "bias" and "fairness" without first addressing whether the system performs adequately.
Many real-world harms stem from AI systems that underperform on their given tasks, they do not achieve an acceptable level of performance. This leads to situations where AI systems not only fail to deliver their promised value but also cause harm. Currently, there is a gap in how we should systematically address this type of AI performance failure. Our paper builds on this discourse by framing the issue as an AI Mismatch, making this discrepancy our central focus.

FATE researchers developed taxonomies of downstream algorithmic harms to help AI practitioners better understand and anticipate potential harms \cite{Shelby2023-ff, barocas2017problem, wang2022measuring, blodgett2020language}. Proactively anticipating harm for AI systems deployed in heterogeneous social contexts is inherently challenging because of the interplay between technologies and social and cultural dynamics \cite{Shelby2023-ff, blodgett2022responsible}. Here, a design approach to AI harm taxonomies can help create actionable resources that help AI practitioners systematically uncover potential sources of harm before committing to building a system. There is emerging interest in the FATE community to employ HCI and design methods to address FATE concerns at earlier stages (e.g., \cite{klumbyte2022critical, shen2022model, Stapleton2022-eb, suresh2022towards}). 
However, concrete processes or actionable guidance for early-stage AI development are still limited \cite{coston2023validity, holstein2019improving, Raji2022-ls}. 
Our work addresses this gap by proposing an approach that supports early AI concept analysis, systematically examining risk factors and revealing critical tradeoffs between risks and benefits.

\section{Method}
Our primary focus was to suggest an approach for examining early-stage AI concepts, specifically aiming to identify potential AI Mismatches. We sought to explore how these mismatches occur between (i) the performance we can realistically expect from AI models and (ii) the performance required for a given task before development begins. As an initial example of how this approach could be applied, we developed a descriptive framework that provides HCI researchers with a valuable lens for identifying and disentangling the complex factors that indicate the likelihood of harm. We envisioned how visualizing the placement of the AI concepts on our framework may help teams more systematically identify which refinements to a design concept might reduce the chance of harm, while maintaining intended benefits. This framework is a preliminary step, demonstrating how our approach can help anticipate scenarios where AI applications might be unintentionally misused.

\subsection{Requirements}
Research shows a strong link between AI failure and tasks requiring high expertise and near-perfect performance, such as recommending treatment for septic shock \cite{yang2019unremarkable, yildirim2023creating}. Focusing on tasks where moderate performance still delivers value could reduce failure risks. To investigate this, we used Yildirim et al.'s \textbf{Task-Expertise x AI-Performance} matrix \cite{yildirim2023creating} to explore whether algorithmic harms are more likely in areas where tasks demand higher expertise or performance. In developing the framework, we were guided by three criteria:
\begin{itemize}
    \item \textbf{Explanatory power}: The framework should help teams articulate and analyze underlying factors in an early-stage AI concept that increase the risks that a concept will cause harm or produce limited value.
    \item \textbf{Analytical leverage}: This framework should support teams in identifying factors that cause the AI to be fundamentally inadequate for the intended task. 
    \item \textbf{Flexibility}: The framework should be flexible enough to accommodate different types of AI applications, industry domains, and contexts.
\end{itemize}

\subsection{Overview of the Process}
We employed an iterative Research through Design (RtD) approach \cite{zimmerman2007research} in our study. Our process consisted of four activities. First, we collected and curated approximately 774 AI application cases. Second, we analyzed these cases using both deductive and inductive approaches. Third, we extracted and synthesized key factors contributing to the mismatch between \textit{required} and \textit{feasible} model performance, which led to potential harm. This process resulted in the development of seven key matrices (see Section $\S$~\ref{sec:matrices}). 

Finally, both our internal research team and external researchers stress-tested these matrices against real-world counterexamples. Our team of five researchers, with expertise in HCI, design, responsible AI, machine learning, psychology, and data science, worked closely with external researchers and practitioners from academia and industry. This included feedback from approximately 20 HCI and AI Fairness academic researchers, 5 data scientists, and 3 designers, ensuring that the most critical factors were thoroughly considered in practice.


\subsubsection{Collecting AI Examples}
We began by reviewing the literature to identify various ways AI can cause harm. 
We then searched three primary sources for real-world AI incidents: 1) the AIAAIC (AI, Algorithmic, and Automation Incidents and Controversies) database, curated and maintained by journalism professionals, 2) the AI Incident database developed by the Partnership on AI, and 3) cases of algorithmic harms discussed in FAccT taxonomies (e.g., \cite{Raji2022-ls, Wang2024-cj, Shelby2023-ff}). Collectively, these three sources offer a rich set of examples where AI has harmed individuals, communities, or society. We also analyzed approximately 64 widely used AI applications, such as Siri, Alexa, fraud detection in finance, and medical imaging in healthcare, as well as a set of 449 industry case studies curated by Evidently AI~\cite{EvidentlyAI}.

Given our focus on harms due to \textbf{\textit{AI Mismatches}}, we excluded cases of deliberate misuse 
such as deepfake pornography or identity theft. Although these issues are important for understanding the broader impact of AI harm, they fall outside the scope of our study. Instead, we focused on cases where the system was used as intended but led to harm due to a fundamental mismatch between the model's required performance and achievable capabilities. 

Our initial collection yielded 
478 cases. To avoid redundancy, we grouped similar incidents—those with comparable causes and outcomes—into single cases, even when they occurred in different organizations. For instance, if two cases involved the use of large language models (LLMs) to produce online articles, both leading to misinformation, we grouped them as one case. As a result, we ended up with 261 AI harm cases. Combined with AI examples outside the AI harm sources, our final set resulted in 774 AI cases. 

\subsubsection{AI Examples Analysis and Extracting Risk Factors}
To begin with, we deductively analyzed 774 AI cases using 
the Task-Expertise x AI-Performance matrix, rating each case based on these two criteria. Next, we inductively analyzed AI cases to unpack these underlying risk factors. Our internal team of five researchers iteratively brainstormed potential factors contributing to AI Mismatches. For example, if a case lacked critical data, we labeled it under "quality of data." At this stage, our team encouraged the generation of as many potential labels as possible for each case, deliberately avoiding premature convergence on a single label to minimize the risk of overlooking any potential causes. Any disagreements regarding specific labels were informally discussed and documented, but final decisions were deferred to the subsequent step of synthesizing these labels. This process was iteratively discussed and cross-checked to ensure that the identified causes accurately represented the reasons for failure. These brainstormed codes were transferred to Post-it notes, ready to be categorized and synthesized.

\begin{table*}[]
\caption{An overview of seven AI Mismatch matrices.}
\resizebox{\textwidth}{!}{
\begin{tabular}{lllll}
\toprule \toprule
\textbf{Matrix Title}         & \textbf{Explanation}                                                                                                                                                                       & \textbf{X-axis}                                                              & \textbf{Y-axis}                                                                                          & \textbf{Section \#}                                         \\ \midrule \midrule
\multicolumn{5}{c}{\color[HTML]{656565}{\textbf{High-level Matrices}}}                                                                                                                                                                                                                                                                                                                                                                                                  \\ \midrule
\rowcolor{gray!15}Required Performance          & \begin{tabular}[c]{@{}l@{}}What is the minimum performance required \\ to create value and prevent harm, and how \\ does this compare to the model's expected \\ performance?\end{tabular} & \begin{tabular}[c]{@{}l@{}}Expected Model\\ Performance\end{tabular}         & \begin{tabular}[c]{@{}l@{}}Minimum Required\\ Performance\end{tabular}                                   & \ref{section:required_performance}       \vspace{0.1cm}\\
Disparate Performance         & \begin{tabular}[c]{@{}l@{}}What level of performance disparity across \\ groups is expected, and how important is it \\ to avoid these disparities for this task?\end{tabular}             & \begin{tabular}[c]{@{}l@{}}Expected Disparity \\ in Performance\end{tabular} & \begin{tabular}[c]{@{}l@{}}Importance of \\ Avoiding Disparities\end{tabular}                            & \ref{section:disparate_performance}       \vspace{0.1cm}\\
\rowcolor{gray!15}Cost of Errors                & \begin{tabular}[c]{@{}l@{}}How does the model's expected performance \\ relate to the severity of potential consequences \\ from errors on this task?\end{tabular}                         & \begin{tabular}[c]{@{}l@{}}Expected Model \\ Performance\end{tabular}        & \begin{tabular}[c]{@{}l@{}}Severity of \\ Error Consequence\end{tabular}                                 & \ref{section:cost_of_errors}            \vspace{0.1cm} \\ \midrule
\multicolumn{5}{c}{\color[HTML]{656565}\textbf{Supporting Matrices to Reflect on Expected Model Performance}}                                                                                                                                                                                                                                                                                                                                                                                 \\ \midrule
\rowcolor{gray!15}Data Quality                  & \begin{tabular}[c]{@{}l@{}}What is the quality of the data we have, \\ and how tolerant is it to low-quality data?\end{tabular}                                                            & Actual Quality of Data                                                       & \begin{tabular}[c]{@{}l@{}}Task Tolerance \\ for Low-Quality Data\end{tabular}                           & \ref{section:data_quality}                                                        \vspace{0.1cm} \\
Model Unobservables           & \begin{tabular}[c]{@{}l@{}}Are there any critical unobserved factors \\ missing from the data? What would be their \\ impact on the model's performance?\end{tabular}                      & \begin{tabular}[c]{@{}l@{}}Importance of \\ Unobserved Factors\end{tabular}  & \begin{tabular}[c]{@{}l@{}}Expected Impact of \\ Unobserved Factors \\ on Model Performance\end{tabular} & \ref{section:model_unobservables}        \vspace{0.1cm}\\ \midrule
\multicolumn{5}{c}{\color[HTML]{656565}\textbf{Supporting Matrices to Reflect on Severity of Error Consequences}}                                                                                                                                                                                                                                                                                                                                                                            \\ \midrule
\rowcolor{gray!15}Expectation of Errors         & \begin{tabular}[c]{@{}l@{}}How tolerant is the task to frequent errors, \\ and how does this align with users' expectations \\ of error frequency?\end{tabular}                            & \begin{tabular}[c]{@{}l@{}}Expected Error \\ Frequency\end{tabular}          & \begin{tabular}[c]{@{}l@{}}Tolerance for Error \\ in Output\end{tabular}                                 & \ref{section:expectation_of_errors}      \vspace{0.1cm}\\
Error Detection \& Mitigation & \begin{tabular}[c]{@{}l@{}}How difficult is it to detect, understand, and \\ mitigate errors for this task, and how much \\ effort is needed to address them?\end{tabular}                 & \begin{tabular}[c]{@{}l@{}}Ease of \\ Error Detection\end{tabular}           & \begin{tabular}[c]{@{}l@{}}Effort Required \\ to Mitigate Errors\end{tabular}                            & \ref{section:error_detection_mitigation} \vspace{0.1cm}\\ \bottomrule \bottomrule
\end{tabular}
}
\label{table:7matrices}
\end{table*}

\subsubsection{Synthesizing Risk Factors}
From the initial set of post-it notes with AI Mismatch factors, we grouped and synthesized the factors that contributed to AI harms, separating them from related but distinct factors. The two co-first authors conducted an initial round of categorizing the factors into emerging themes. When conflicting factors were grouped within a single concept, they carefully deliberated to identify the most appropriate ones to include. Following this initial categorization, the internal research team (n=5) engaged in a series of weekly discussions over 12 months. Between meetings, team members shared opinions, concerns, and suggestions asynchronously using collaborative web applications. These comments were consolidated by the first authors and incorporated into updated drafts for discussion in subsequent meetings. Throughout this iterative process, some factors were decoupled or synthesized to strike a balance between achieving a suitable level of abstraction and ensuring conceptually distinct categories.
For instance, we decoupled users’ expectations of errors from the need for error detection and mitigation. Some AI examples, like creative AI tools or smartwatch step counters, showed that users expect errors but have a higher tolerance for them. These users are less concerned with accuracy and more interested in patterns or inspiration. On the other hand, in cases like AI for financial fraud detection, users expect errors but also require accurate error detection and mitigation. This highlighted that, beyond error expectation, the tolerance for error in AI output is a distinct factor. This distinct factor was documented and added to our set of post-it notes. 

We then explored additional risk factors that could influence the initial risk factor, placing these on the horizontal and vertical axes of a matrix to identify high-risk areas. 
For instance, the severity of errors correlates with 
the prevalence of errors:  a higher prevalence of low-severity errors could be acceptable (e.g., language translation), whereas high-severity errors should be rare (e.g., industrial automation). An AI system with frequent severe errors is problematic. 
Similarly, we recognized that ease of mitigation is related to ease of detection for high-severity errors; errors that are hard to detect and mitigate would render the AI concept infeasible (e.g., LLMs for medical diagnosis \cite{hager2024evaluation}). 
The interplay between these risk factors  — severity, expectation, detection, and mitigation of errors — indicates the need to systematically unpack them. Drawing upon the examples of creative AI tools and the smartwatch step counter, since the severity of individual errors is negligible and users expect these errors to occur with minimal impact on value creation, it is not necessary to be able to detect these errors or mitigate them.

We recognized that it was hard to assess the riskiness of an AI concept because AI systems are often introduced in high-risk scenarios where the AI concept may unexpectedly add more risks, consequently blurring the difference between the risk posed by AI versus the risks present in the scenario or environment. For example, in lane departure prediction, drivers may swerve due to fatigue or road conditions, but an AI with frequent errors (i.e., false positives) could distract drivers and increase the risk of accidents. Conversely, isolated cases of harm, such as following faulty GPS directions off a collapsed bridge \cite{AIAAIC_driver} or a pedestrian walking down the highway \cite{AIAAIC_pedestrian}, are rare and don’t usually reduce the overall value of AI. These incidents may also result from external factors such as missing barricades or poor judgment. In sum, we assessed the riskiness of AI concepts by asking, "\textit{Does this AI concept consistently or frequently add more risk to the baseline?} (i.e., the scenario or environment before the use of AI)."

\vspace{0.2cm}
In this paper, we adopt \textbf{a human-centric definition of model performance} \cite{parasuraman2000model, yang2019unremarkable, lai2021towards}. We define model performance in terms of the \textbf{model's ability to perform a task that fulfills a human need}. This formulation is important because AI concepts may appear to exhibit excellent performance per traditional AI evaluation methods and metrics, but exhibit much lower performance on the real-world task that they are actually meant to support. For example, consider an AI predictive model that is intended to provide decision-support to social workers. The model might show high predictive accuracy but fail to support social workers’ decisions if its predictions are irrelevant to their decision-making needs \cite{saxena2024algorithmic, Kawakami2022-tx}.

As a result of the iterative synthesis process, we consolidated the AI Mismatch factors into seven key factors that indicate a high likelihood of harm. The resulting seven matrices were iteratively refined through a series of ongoing discussions among the authors and external researchers. Next, we explored various ways to visualize the relationships between these factors. Since we found that each factor was best represented in two dimensions, allowing us to map a plane indicating high-risk areas, we ultimately chose a 3x3 matrix for our visualization.

\begin{figure*}[t!]
    \centering
    \includegraphics[width=\linewidth]{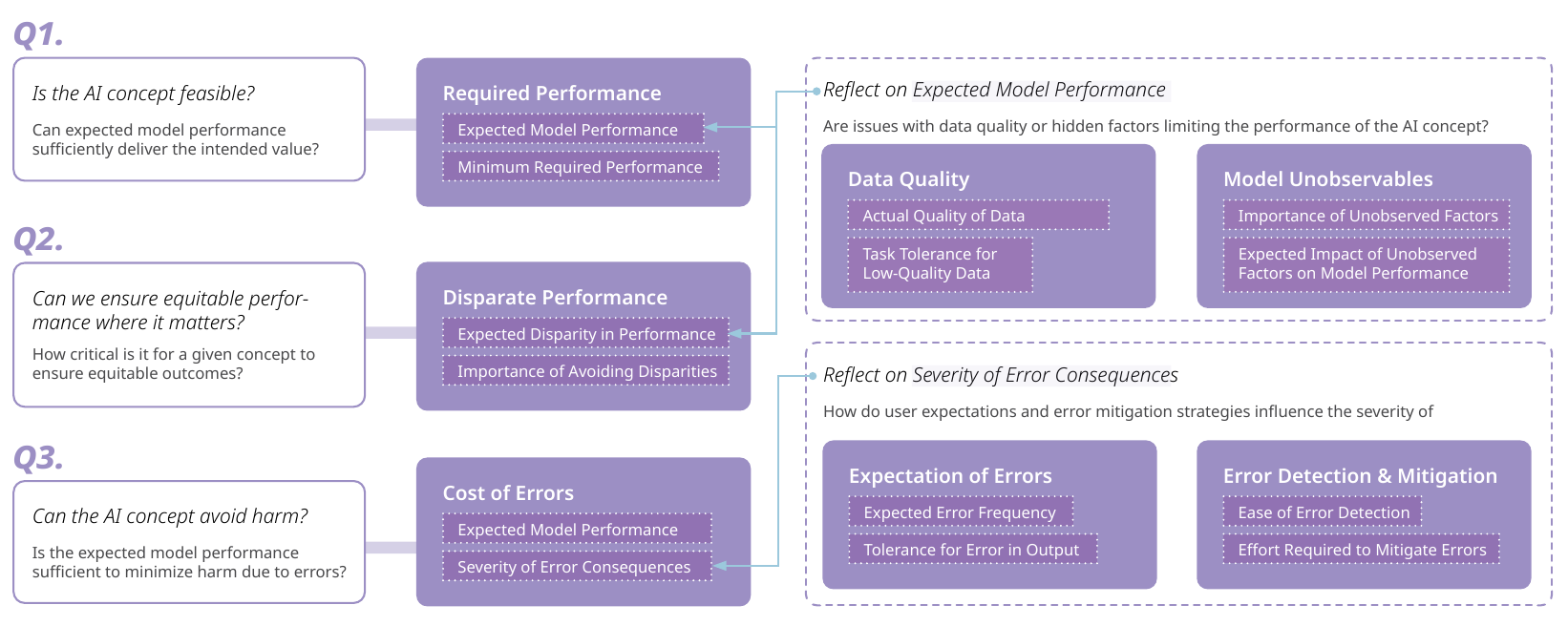}
    \caption{Motivations and Interconnections of the Seven AI Mismatch Matrices}
    \label{fig:matrices-relationship}
    \Description[Seven AI Mismatch Interconnectedness Overview]{This visual summarizes the relationship between seven AI Mismatch Matrices. On the left, there are three motivating questions that guide readers to one of three core matrices. Then these core matrices are assisted by supporting matrices with prompting questions on the right.}
\end{figure*}

\section{Overview of the Matrices}~\label{sec:overview}
Our synthesis identified seven key factors that impact AI Mismatches: (i) Data Quality, (ii) Model Unobservables, (iii) Expected Performance, (iv) Cost of Errors, (v) Disparate Performance, (vi) Expectation of Errors, and (vii) Error Detection and Mitigation. To help communicate, we visualized the factors as 3x3 matrices with two axes that represent different variables of interest, focusing on areas of potential risk. 
Table~\ref{table:7matrices} offers a comprehensive summary, including titles, explanations, axes, and section references for further exploration.
While Section $\S$~\ref{sec:matrices} provides detailed visualizations and explanations for each matrix, this section provides a single, integrated view of the framework and explains how the matrices connect, inform one another, and guide decision-making.

Among seven AI Mismatch matrices, the framework consists of three high-level matrices aimed at helping teams reflect on potential mismatches between expected model performance and expected real-world value and risks. These matrices help determine whether a concept’s expected performance is sufficient to deliver the intended value (i.e., Required Performance Matrix), whether there are expected performance disparities that matter (i.e., Disparate Performance Matrix), and whether its expected performance is sufficient to minimize harm due to errors (i.e., Cost of Errors Matrix). In addition to these three core matrices, we present four examples of lower-level matrices that can help inform these higher-level judgments, such as assessments of expected performance or expected consequences of errors. Figure~\ref{fig:matrices-relationship} presents a simplified diagram illustrating how these matrices interconnect and support one another in guiding decision-making. The left side of the figure presents three motivating questions that lead to the three core matrices. On the right, four supporting lower-level matrices with two prompting questions, inform one axis of the core matrices. Additional details are provided in a step-by-step workflow (Section $\S$~\ref{sec:workflow}) and a one-page flyer (Section $\S$~\ref{sec:flyer}) in the appendix.

\subsection{Performance Mismatch Matrices}
\noindent These are the core matrices that highlight the critical high-level judgments teams must make.

\vspace{0.3cm}
\textbf{Required Performance Matrix} (\textit{Expected vs. Required Performance}) -- This matrix asks, “Is the expected model performance sufficient to deliver the intended value?” It encourages teams to move beyond technical metrics to consider what is the minimum level of performance required to create actual value for people or organizations. Then it asks teams to reflect on what level of performance is likely to be achievable in practice, to anticipate potential performance mismatches.

\textbf{Disparate Performance Matrix} (\textit{Expected Disparity in Performance vs. Importance of Avoiding Disparities}). This matrix introduces an equity lens and asks, “Can we ensure equitable performance where it matters?” Even if a model meets minimum performance for most users, are there significant performance gaps across different demographic or user groups? How critical is it for a given AI concept to ensure equitable outcomes?

\vspace{0.1cm}
\textbf{Cost of Errors Matrix} (\textit{Expected Performance vs. Severity of Error Consequences}) -- This matrix complements the Required Performance matrix by looking beyond benefits and asking, “Is the expected model performance sufficient to minimize harm due to errors?”. It is not enough for a model to perform well on average; we must understand what happens when it fails. Are these failures merely inconvenient, or could they cause real harm?

\vspace{0.1cm}
These three matrices address technical feasibility, practical value, risk tolerance, and equity considerations to support reflection on potential mismatches in AI concepts. They offer high-level checks that help practitioners understand whether the AI concept meets the thresholds of real-world value and acceptable risk.

\vspace{-0.2cm}
\subsection{Examples of Lower-Level, Supporting Matrices}
\noindent Below, we present examples of lower-level matrices that can help support the high-level judgments required by the core matrices (e.g., assessments of expected performance).

\subsubsection{Scaffolding Reflection on Expected Performance} The \textbf{Data Quality Matrix} asks teams to reflect upon the extent to which limitations of the underlying data may impact model performance in ways that matter. As a second example, the \textbf{Model Unobservables Matrix} encourages reflection on what information the model may be missing, and the implications of these information gaps for model performance. Taken together, these two matrices support reflection on what level of performance can be realistically expected.

\subsubsection{Scaffolding Reflection on the Severity of Error Consequences} The \textbf{Expectation of Errors Matrix} encourages reflection on user expectations, which can amplify or reduce the perceived severity of errors. If users are likely to expect errors, the impact might feel less severe; if they expect near-perfection, even mild errors can be perceived as catastrophic. As a second example, the \textbf{Error Detection \& Mitigation Matrix} draws attention to the fact that knowing how easily errors can be detected and fixed can change how severe they feel operationally. On the other hand, difficult-to-detect errors, even if rare, have the potential to lead to more severe consequences. Taken together, these two matrices can help support reflection on an AI concept’s potential for real-world harm that may be due to errors.

In summary, \textbf{Required Performance}, \textbf{Cost of Errors}, and \textbf{Disparate Performance Matrices} are the central “decision dashboards” for high-level viability judgments, while lower-level matrices like the examples above may act as diagnostic or "drill down" tools and support the use of core matrices. This layered approach may allow teams to move from a broad, human-centered perspective (i.e., core matrices) down to the root causes and potential opportunities for idea refinement (supporting matrices).

\section{AI Mismatch Matrices}\label{sec:matrices}

The previous section (Section $\S$~\ref{sec:overview}) provided an overview of the seven matrices within the AI Mismatch Matrices framework, highlighting their relationships and interdependencies. Building on that foundation, this section delves deeper into each matrix, offering a detailed explanation of their structure and purpose. These matrices are: (i) Data Quality, (ii) Model Unobservables, (iii) Expected Performance, (iv) Cost of Errors, (v) Disparate Performance, (vi) Expectation of Errors, and (vii) Error Detection and Mitigation. The AI Mismatch approach illustrates how conceptually flawed AI systems can inadvertently cause harm. Our framework uses 3x3 matrices to visualize and highlight these areas of risk. Each matrix is structured with two axes, representing different and often opposing variables of interest, to clarify the underlying tensions or trade-offs. This detailed exploration aims to illuminate how each matrix contributes to identifying and addressing potential mismatches in AI systems.

The matrices use a color gradient to indicate risk levels
: red represents the highest risk, pink portrays an intermediate risk, yellow indicates moderate risk, and uncolored areas represent the low risk. R
ed zones highlight 
significant gap 
between required and actual performance, posing 
substantial risk. We encourage teams to aim for 
AI concepts 
in low-risk, uncolored zones wherever possible. However, if work in higher-risk zones is unavoidable, careful consideration should be given to mitigating potential negative impacts. 

These matrices are meant to foster interdisciplinary collaboration, bringing together teams with technical, ethical, and design expertise to make informed judgments about AI concepts. By collectively assessing where an AI concept falls within the matrix, teams can identify and discuss potential concerns or uncertainties early in the development process. 
This initial assessment is a dynamic process, evolving over time, and aims to highlight AI Mismatches early, allowing teams to address issues before development advances. The hypothetical nature of this process highlights the importance of interdisciplinary expertise in providing comprehensive evaluations and ensuring AI Mismatches are detected as early as possible.

\subsection{Performance Mismatch Matrices}
In the following subsections, we provide a walkthrough of our three core matrices.

\vspace{0.3cm}
\subsubsection{{Required Performance Matrix}}~\label{section:required_performance}
\begin{figure}
    \centering
    \includegraphics[width=0.8\linewidth]{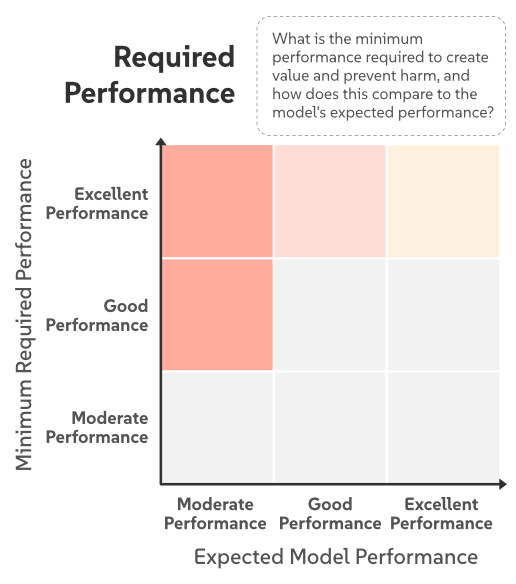}
    \caption{The 
    {Required} Performance matrix}
    \label{fig:3-expected-performance}
    \Description[Required Performance matrix]{Matrix showing the relationship between the minimum required performance (Moderate to Excellent) and expected model performance (Moderate to Excellent), with darker colors representing higher risks.}
\end{figure}

The Required Performance matrix draws attention to the contrast between the AI model's expected performance under current constraints (i.e., data quality and model unobservables) and the minimum performance needed to create value. {Figure~\ref{fig:3-expected-performance} represents the matrix as a 3x3 grid. The 
Expected Model Performance is plotted on the X-axis, and the Minimum Required Performance is on the Y-axis.} \textbf{Minimum Required Performance to Create Value} defines the baseline at which the AI must operate to be effective, avoid harm, and provide value. For example, AI used to help content moderation on social media typically requires moderate performance. False positives or missed content can be reviewed by human moderators. However, in K-12 online learning, excellent performance is essential to protect children from harmful content. Small errors that expose children at school to harmful content can harm their well-being and damage the school’s reputation. \textbf{Expected Model Performance} reflects how well the AI is expected to perform given current technological constraints. For instance, GitHub Copilot’s coding assistance performs well due to the structured nature of code, while mobile device autocorrect performs only moderately well due to the complexity of language and the messiness of texting communication.

When the expected model performance on a given task is lower than the required performance, it raises an important contradiction that makes the AI concept infeasible. For instance, facial scanning for biometric security on personal devices performs exceptionally well because the task is straightforward: matching a face to saved data, similar to showing an ID to a bartender. In contrast, facial recognition for surveillance is much more complex, involving unobserved factors (e.g., lighting, weather, inconsistent angles of image capture) that lead to a higher error rate and lower model performance. It is akin to asking a bartender to identify a person from numerous profile pictures. Consequently, facial recognition used by police departments has led to wrongful arrests and caused significant harm to communities \cite{Buolamwini2023-xl}.

\vspace{0.1cm}
\textbf{\textit{Rationale.}}	Through our case studies, we decoupled the goals of AI applications from their actual model performance, uncovering the discrepancy highlighted by this matrix. In the deductive analysis, we observed that many AI harm cases required high performance to create value. However, when we inductively analyzed the causes of harm, we found that these AI systems often only achieved moderate or good performance. This performance gap is often due to factors like poor data quality, unrealistic problem formulation, or missing unobserved factors that are critical to task success.

\begin{figure}
    \centering
    \includegraphics[width=0.8\linewidth]{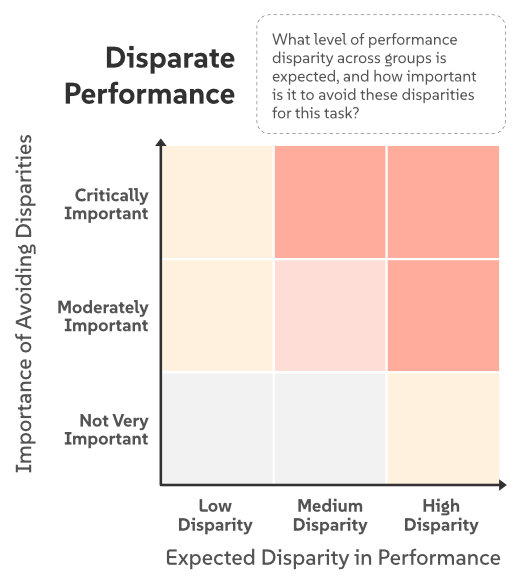}
    \caption{The Disparate Performance matrix.}
    \label{fig:5-disparate-performance}
    \Description[Disparate Performance matrix]{Matrix showing the relationship between expected disparity in performance (Low to High) and the importance of avoiding disparities (Not very important to Critically important), with darker colors representing higher risk.}
\end{figure}

\vspace{0.2cm}
\subsubsection{Disparate Performance Matrix}~\label{section:disparate_performance}
The Disparate Performance matrix highlights performance disparities and their ethical implications. {Figure~\ref{fig:5-disparate-performance} represents the matrix as a 3x3 grid. The Expected Disparity in Performance is plotted on the X-axis, and the Importance of Avoiding Disparities is on the Y-axis.} 
\textbf{Expected Disparity in Performance} refers to how well AI models perform across different groups (e.g., racial, socio-economic, gender). High disparity means the AI is expected to perform significantly better for some groups than others, often due to biased data or structural issues. Low disparity indicates that performance is consistent across groups, with minimal variation, typically achieved through balanced or bias-reducing training data. For instance, facial recognition AI may exhibit low disparity in controlled settings like office security, but disparities increase in public spaces like airports, where environmental variability impacts performance. \textbf{Importance of Avoiding Disparities} assesses how critical it is to avoid performance gaps, based on the ethical, social, or financial consequences of unequal outcomes. In less critical contexts, disparities might be considered "not very important," leading only to minor user dissatisfaction. However, in high-stakes settings, avoiding disparities becomes crucial, as unequal AI performance can reinforce bias and discrimination, raising significant ethical and legal concerns.
 
This matrix is crucial in early AI development, helping teams assess and mitigate performance disparities. For instance, in language translation (e.g., Google Translate), disparate performance is often anticipated when translating between less commonly used languages due to the limited availability of parallel text for training \cite{schaffner2016parallel}. On the other hand, in clinical AI tools, disparate performance may occur because different demographic groups have varying base rates for disease prevalence \cite{Stanley2022-ad}. However, racial disparities in the healthcare system confound these base rates \cite{obermeyer2019dissecting} leading to a high ethical cost of disparate performance. This further helps assess the effort needed to prevent or mitigate disparities, and whether that effort creates enough value that outweighs the risks. For instance, Holstein et al. \cite{holstein2019improving} found that image recognition systems that mislabeled female doctors as nurses could use the umbrella term of ‘medical professionals’ and avoid using ‘doctors’ and ‘nurses’ as labels; a simple redesign that mitigates disparate classification. 

\vspace{0.1cm}
\textbf{\textit{Rationale.}}	Disparate performance from data imbalances was observed in AI systems such as facial recognition and speech recognition, where some demographic groups might be underrepresented in the training data. Disparate performance due to differing base rates for different demographic groups means that there are underlying characteristics (e.g., socioeconomic factors) due to which the frequency of observed outcomes varies for the different groups. This has been observed in AI systems such as predictive policing \cite{haque2024we}, loan default prediction \cite{garcia2024algorithmic}, and AI hiring \cite{roemmich2023values}, among others where AI predictions reflect these underlying disparities and present an inherent pitfall because it is not possible to achieve fair predictions for different groups when the base rates vary \cite{Chouldechova2017-aj, Kleinberg2016-fd, Wang2024-cj}.

\aptLtoX{\begin{figure}[htbp]
    \centering
        \includegraphics[width=0.8\columnwidth]{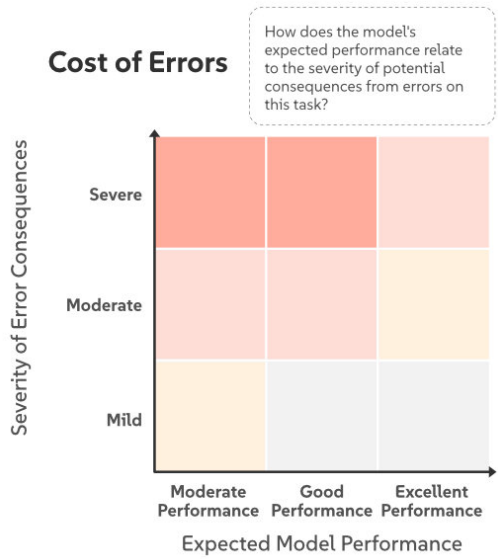}
        \caption{The Cost of Errors matrix.}
        \label{fig:4-cost-of-errors}
        \Description[Cost of Errors matrix]{Matrix showing the relationship between expected model performance (Moderate to Excellent) and severity of error consequences (Mild to Severe), with darker colors representing higher risk.}
    \end{figure}
    \begin{figure}
        \centering
        \includegraphics[width=0.8\columnwidth]{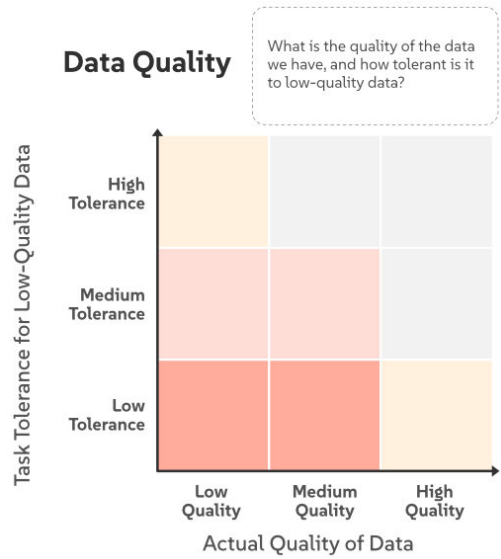}
        \caption{The Data Quality matrix.}
        \label{fig:1-data-quality}
        \Description[Data Quality three by three Matrix]{Matrix showing the relationship between actual data quality (Low to High) and task tolerance for low-quality data (Low to High), with darker colors indicating lower tolerance and quality.}
    \end{figure}}{\begin{figure*}[htbp]
    \centering
    \begin{minipage}{\columnwidth}
        \centering
        \includegraphics[width=0.8\linewidth]{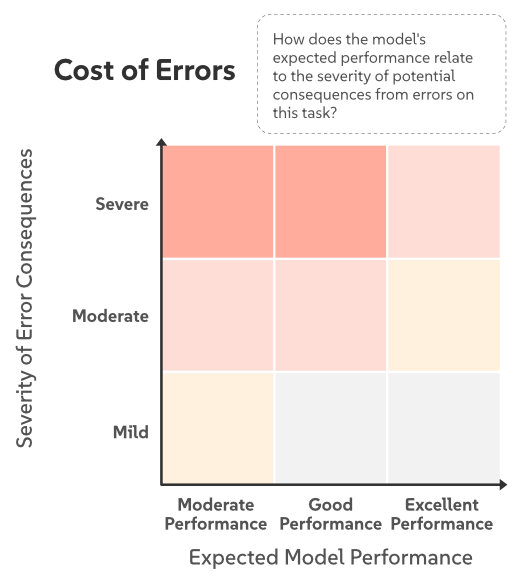}
        \caption{The Cost of Errors matrix.}
        \label{fig:4-cost-of-errors}
        \Description[Cost of Errors matrix]{Matrix showing the relationship between expected model performance (Moderate to Excellent) and severity of error consequences (Mild to Severe), with darker colors representing higher risk.}
    \end{minipage}
    \begin{minipage}{\columnwidth}
        \centering
        \includegraphics[width=0.8\linewidth]{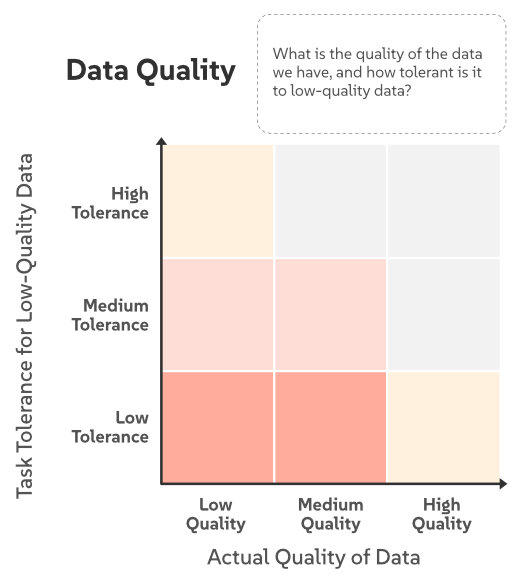}
        \caption{The Data Quality matrix.}
        \label{fig:1-data-quality}
        \Description[Data Quality three by three Matrix]{Matrix showing the relationship between actual data quality (Low to High) and task tolerance for low-quality data (Low to High), with darker colors indicating lower tolerance and quality.}
    \end{minipage}
\end{figure*}}

\vspace{0.3cm}
\subsubsection{Cost of Errors Matrix}~\label{section:cost_of_errors}
The Cost of Errors matrix offers an alternative frame for thinking about the expected model performance matrix by considering the severity and prevalence of errors. {Figure~\ref{fig:4-cost-of-errors} represents the matrix as a 3x3 grid. The Expected Model Performance is plotted on the X-axis, and the Severity of Error Consequences is on the Y-axis.} Similar to the first three matrices, the horizontal axis represents \textbf{Expected Model Performance}. However, for this matrix, we use the prevalence of errors to signify the inverse direction for model performance: higher performance means fewer errors, and vice versa. \textbf{Severity of Error Consequences} on the vertical axis measures how serious the consequences would be if the AI makes an error. Here, we see severity levels as highly context-dependent, and the interpretation of severity should be discussed between relevant stakeholders. For example, 
severe consequences could include cases that directly harm people, organizations, or systems and are difficult to mitigate, such as high-speed crashes in autonomous driving. In contrast, mild consequences lead to minor inconveniences, dissatisfaction, or inefficiencies, which are easily recovered from. For example, if Spotify recommends an unwanted song, the user can simply skip it, resulting in minimal frustration. 

The intersection of severity and prevalence of errors is an important consideration because rare but severe errors or prevalent but mild errors might be acceptable scenarios for AI concepts. However, errors that are both severe and prevalent would carry high ethical and/or financial risks, making the AI concept harmful (e.g., IBM Watson for Oncology \cite{Faheem2023-ky}). The matrix draws attention to the necessary tradeoffs (i.e., a cost-benefit analysis) to co-create value. For instance, warehouses created controlled environments for robots, including well-defined floor plans and storage locations, and limited pedestrian traffic during hours of operation to significantly lower the prevalence of errors. Investment banks, on the other hand, have invested significant resources in risk mitigation to lower the severity of errors as a result of algorithmic trading \cite{Addy2024-lv}. 

\vspace{0.1cm}
\textbf{\textit{Rationale.}}	Through our case studies, we found that many AI failures were not just a result of low expected model performance, but also due to the underestimation of how severe and frequent errors could be in practice. For example, in healthcare applications like IBM Watson for Oncology, errors in diagnostic suggestions, while rare, carry severe consequences that could affect patient outcomes \cite{Faheem2023-ky}. Cases like this illuminated the need for a matrix that not only captures expected performance but also considers the broader consequences of errors, thus offering a more comprehensive understanding of the risks associated with AI deployment. 

\vspace{-0.1cm}
\subsection{Supporting Matrices}
Below, we walk through four examples of lower-level matrices that can help support the higher-level judgments required by the core matrices described above.

\aptLtoX{\begin{figure}[t!]
        \centering
        \includegraphics[width=0.8\columnwidth]{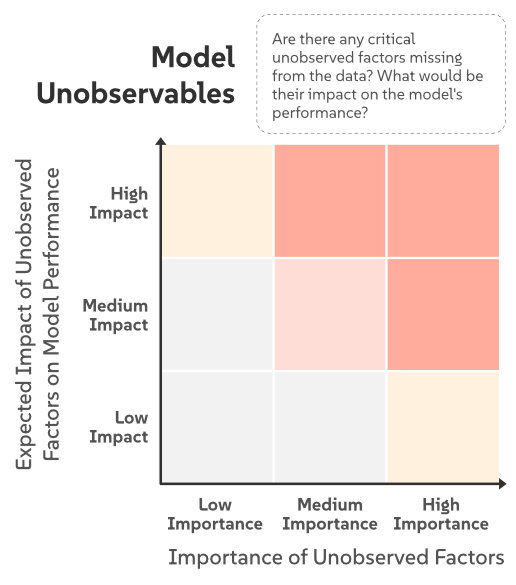}
        \caption{The Model Unobservables matrix.}
        \label{fig:2-model-unobservables}
        \Description[Model Unobservables matrix]{Three by three matrix illustrating the relationship between the importance of unobserved factors (Low to High) and their expected impact on model performance (Low to High), with darker shades indicating higher risk.}
    \end{figure}
    \begin{figure}
        \centering
        \includegraphics[width=0.8\columnwidth]{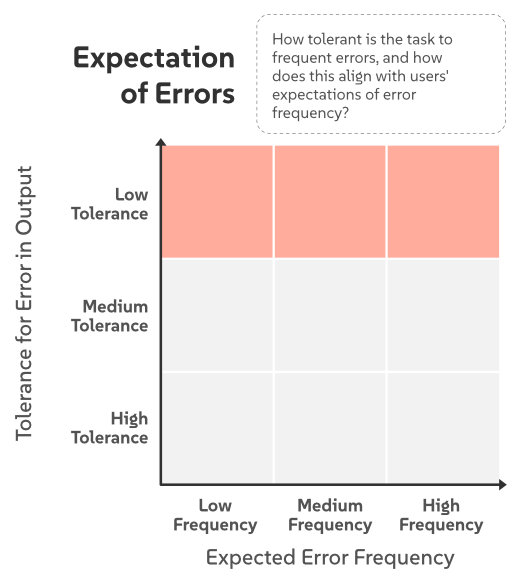}
        \caption{The Expectation of Errors matrix.}
        \label{fig:6-expectation-of-errors}
        \Description[Expectation of Errors matrix]{Matrix showing the relationship between expected error frequency (Low to High) and tolerance for error in AI output (Low to High), with darker colors representing higher risk.}
\end{figure}}{\begin{figure*}[htbp]
    \centering
    \begin{minipage}{\columnwidth}
        \centering
        \includegraphics[width=0.8\linewidth]{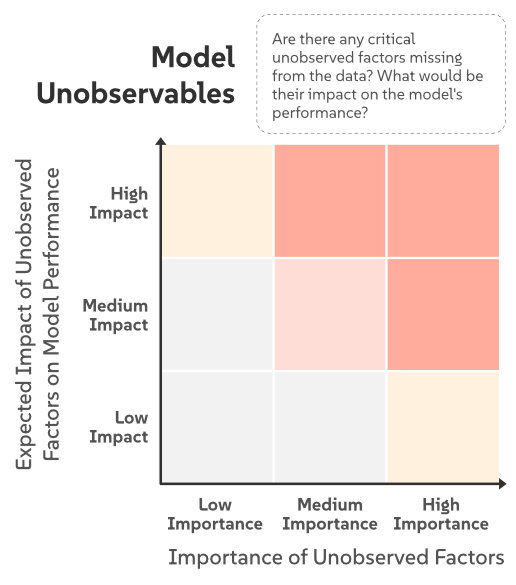}
        \caption{The Model Unobservables matrix.}
        \label{fig:2-model-unobservables}
        \Description[Model Unobservables matrix]{Three by three matrix illustrating the relationship between the importance of unobserved factors (Low to High) and their expected impact on model performance (Low to High), with darker shades indicating higher risk.}
    \end{minipage}
    \begin{minipage}{\columnwidth}
        \centering
        \includegraphics[width=0.8\linewidth]{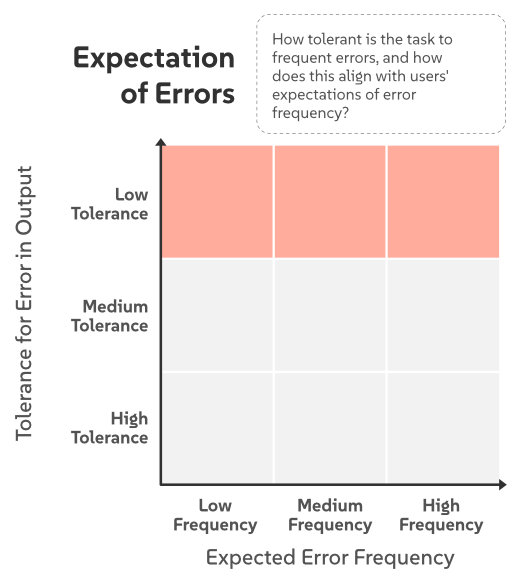}
        \caption{The Expectation of Errors matrix.}
        \label{fig:6-expectation-of-errors}
        \Description[Expectation of Errors matrix]{Matrix showing the relationship between expected error frequency (Low to High) and tolerance for error in AI output (Low to High), with darker colors representing higher risk.}
    \end{minipage}
\end{figure*}}

\vspace{0.2cm}
\subsubsection{Data Quality Matrix}~\label{section:data_quality}
The Data Quality matrix highlights potential mismatches between data quality (completeness, accuracy, consistency, timeliness, validity) and an AI design's tolerance for lower-quality data. {Figure~\ref{fig:1-data-quality} represents the matrix as a 3x3 grid. The Actual Quality of Data is plotted on the X-axis, and the Task Tolerance for Low-Quality Data is on the Y-axis.} \textbf{Actual Quality of Data} refers to the quality of available data to build the AI concept. Low-quality data is often poorly labeled, incomplete, outdated, or noisy—such as a user's music history, where skips, incomplete plays, and ambiguous preferences introduce noise. High-quality data is structured, consistent, and standardized, as seen in medical imaging (e.g., MRI or CT scans), which is precise and standardized for clinical use. \textbf{Task Tolerance for Low-Quality Data} refers to how much the AI can handle lower-quality data without significant performance degradation. Low tolerance means even small data inconsistencies or inaccuracies significantly impact performance, especially in rigid models or high-stakes contexts. For example, manufacturing quality control AI, which monitors for defects on a production line, requires consistently high-quality data to detect deviations. 
In contrast, AI for disaster prediction (e.g., floods or wildfires) can tolerate some missing or noisy data, such as incomplete weather reports, while still making reasonable predictions meant to impact planning.

It is crucial to address this conceptual mismatch early on, as new data in adaptive systems can harm performance (e.g., Tay chatbot \cite{davis2016ai}). If high variation in data quality is expected, the AI design should be refined to better tolerate lower-quality data. For instance, AI for medical diagnosis requires rigorous data cleaning and validation, while media recommenders such as {Spotify and Netflix}, can tolerate lower-quality data, allowing more focus on improving modeling. If high-quality data isn't achievable, the AI concept must be revised to align with the system's data tolerance.

\vspace{0.1cm} 
\textbf{\textit{Rationale.}}	 Our case study analysis revealed many AI concepts where poor data quality undermined the AI’s inferences. High-risk applications like AFST \cite{Kawakami2022-tx}, COMPAS \cite{rudin2020age}, and FraudDetect \cite{epic_report_2023} require high-quality data to support critical decisions (e.g., risk of child maltreatment, recidivism, or welfare fraud). However, public sector data often lacks this quality, and even post hoc improvements may not address fairness concerns \cite{Allard2018-hu, Logan2016-of, Botsis2010-ac}. Additionally, as previously noted, de-biasing methods can amplify harmful biases if the data is severely flawed \cite{Cooper2021-ft, dutta2020there, kallus2018residual}. Many representational and allocative harms can be traced to data quality issues \cite{Shelby2023-ff, crawford_2017}.

\vspace{0.3cm}
\subsubsection{Model Unobservables Matrix}~\label{section:model_unobservables}
The Model Unobservables matrix draws attention to the role of unobserved factors—variables that influence task performance but are not captured in the model. {Figure~\ref{fig:2-model-unobservables} represents the matrix as a 3x3 grid. The Importance of Unobserved Factors is plotted on the X-axis, and the Expected Impact of Unobserved Factors on Model Performance is on the Y-axis.} Model unobservables affect the outcome but are not directly measured or included in the model as features and can lead to biased or inaccurate predictions \cite{kleinberg2018human, ludwig2021fragile, guerdan2023ground}. For instance, stress and sleep quality affect a student's test score but are not included in official performance data, potentially resulting in incomplete or skewed predictions. Understanding these missing predictors is crucial, particularly in high-stakes applications like healthcare or autonomous systems, where errors can have severe consequences.

\vspace{0.1cm}
\textbf{Importance of Unobserved Factors} axis refers to how the absence of missing variables could lead to serious model errors or oversights. For example, a patient's genetic history or environmental exposure is often unrecorded but essential for diagnosis. In contrast, for media recommenders, unobserved factors like users' temporary preferences are less critical, as general-purpose algorithms still provide reasonable suggestions. The \textbf{Expected Impact of Unobserved Factors on Model Performance} axis reflects how their absence affects outcomes. A high impact suggests that missing these factors significantly impairs model performance. For example, in autonomous driving, unobserved environmental hazards can result in unsafe decisions. Low impact indicates that missing unobserved factors have little effect, such as in predicting long-term energy consumption, where occasional household spikes have minimal influence on predictions. 

\vspace{0.1cm}
\textbf{\textit{Rationale.}} We found several AI concepts across domains (e.g., criminal justice \cite{rudin2020age}, healthcare \cite{obermeyer2019dissecting}, policing \cite{rezende2020facial}, hiring \cite{brown2023hiring}, and credit learning \cite{leal2024algorithms}) where unobserved factors such as socioeconomic conditions, systemic biases, or historical inequalities led to biased predictions and downstream harm. Formulating AI tasks requires data scientists to convert high-level objectives into computational problems where they focus on quantifiable factors while inadvertently abstracting out unobserved contextual factors that affect real-world tasks \cite{passi2019problem, selbst2019fairness, saxena2023rethinking, Kawakami2022-ez}. For instance, a judge considers a defendant’s demeanor and attitude during bail sentencing, and a physician considers a patient’s visible symptoms that a model does not capture \cite{guerdan2023ground}. Missing critical unobserved factors can significantly impact model performance, leading to high uncertainty and potential harm. This gap often reveals discrepancies between prediction targets and actual constructs, with fairness and bias concerns frequently linked to these mismatches \cite{Jacobs2021-of, guerdan2023ground}.

\vspace{0.3cm}
\subsubsection{Expectation of Errors Matrix}~\label{section:expectation_of_errors}
The Expectation of Errors matrix captures the user's expectations about AI error frequency and their tolerance for errors in AI output. {Figure~\ref{fig:6-expectation-of-errors} represents the matrix as a 3x3 grid. The Expected Error Frequency is plotted on the X-axis, and the Tolerance for Error in Output is on the Y-axis.} \textbf{Expected Error Frequency} shows how often users expect the AI to make mistakes. When users expect a high frequency of error, they brace for AI’s mistakes. In contrast, low error expectations suggest users believe the system performs correctly most of the time. For example, users expect less frequent errors while navigating on well-mapped roads on Google Maps, while they usually expect more errors when they are using it in less-mapped or off-road areas. \textbf{Tolerance for Error in Output} measures how accepting the users or the task environment are of these mistakes or inaccuracies in AI's output. A high tolerance for ambiguity means users are more comfortable with vague outputs, whereas a low tolerance demands precise, clear outputs or explanations. For instance, for creative AI tools, users expect frequent errors and may even find unexpected output (errors or ambiguous results) helpful for their creative process \cite{Zhou2023-cf}. 

\vspace{0.1cm}
The matrix helps identify where AI concepts are most susceptible to causing harm, especially when tolerance for errors or ambiguity is low. Understanding this balance is crucial for building appropriate trust and reliance in AI systems. Without it, professional practices can suffer. For example, the EPIC sepsis prediction model failed partly due to a high false positive error rate, which led to alert fatigue for practitioners \cite{Wong2021-zv}. 

\vspace{0.2cm}
\textbf{\textit{Rationale.}} There is an irreducible degree of uncertainty associated with any AI inference \cite{de2020case, saxena2021framework}, which inevitably leads to incorrect decisions. Therefore, it is essential that users have a clear understanding of when and how often they should expect errors. Through our analysis, we found several AI concepts where the users were not provided with enough (if any) information about error conditions with several concepts purposely withholding this information to prevent `gaming behavior' \cite{Kawakami2022-ez}.

\begin{figure}[t!]
    \centering
    \includegraphics[width=0.8\linewidth]{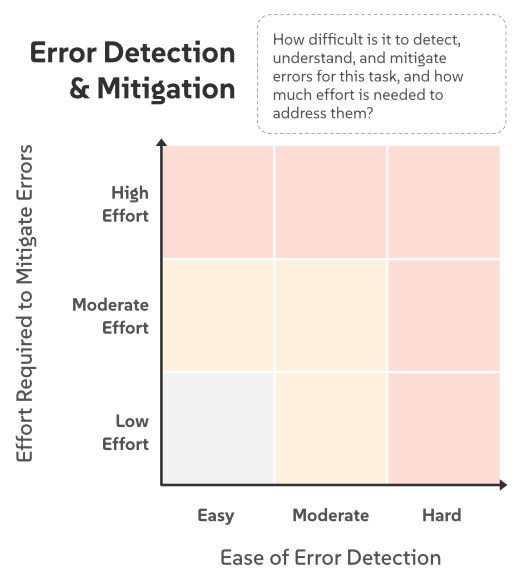}
    \caption{The Error Detection \& Mitigation matrix.}
    \label{fig:7-error-detection-mitigation}
    \Description[Error Detection \& Mitigation matrix]{Matrix showing the relationship between ease of error detection (Easy to Hard) and effort required to mitigate errors (Low to High), with darker colors representing higher risk.}
\end{figure}

\vspace{0.3cm}
\subsubsection{Error Detection \& Mitigation Matrix }~\label{section:error_detection_mitigation}
In the Error Detection \& Mitigation matrix, the two axes evaluate the ease of detecting errors in AI systems and the effort required to mitigate those errors once they are identified. {Figure~\ref{fig:7-error-detection-mitigation} represents the matrix as a 3x3 grid. The Ease of Error Detection is plotted on the X-axis, and the Effort Required to Mitigate Errors is on the Y-axis.} \textbf{Ease of Error Detection} axis measures how readily users or systems can identify AI errors. Errors are easy to detect when they are visible and straightforward, requiring no specialized knowledge—like spotting spelling mistakes in a word processor. Detection becomes difficult when errors require expertise, deep investigation, or time to become apparent, such as miscounted steps in fitness tracking data or incorrect heart rate measurements. \textbf{Effort Required to Mitigate Errors} axis reflects the time, resources, or expertise needed to address errors once identified. Low-effort mitigation involves quick, simple fixes with minimal resources, such as ignoring irrelevant recommendations on {Netflix}. High-effort mitigation requires substantial resources, expertise, or system overhauls, like retraining models in financial fraud detection systems or updating algorithms. 

\vspace{0.2cm}
If the severity of errors is high and they are hard to detect, then the AI concept needs to be scoped within a comprehensive risk mitigation framework. AI-driven stock trading offers an interesting case-in-point where the financial sector has invested significant resources toward mitigating the impact of errors such that benefits outweigh the harms \cite{Glantz2014-zr}. On the other hand, if the severity of errors is low and they are easy to detect, mitigation may not be necessary (e.g., creative AI tools). The matrix can help prioritize AI concepts based on their position on the matrix by assessing the tradeoffs between effort and resources required to detect and mitigate errors versus the purported value creation. Strategic planning around this can help avoid downstream harm to professional practice. 

\vspace{0.2cm}
\textbf{\textit{Rationale.}} We discovered AI concepts that initially promised value but resulted in severe or prevalent errors, making detection and mitigation labor-intensive. For AI concepts to truly create value, errors must be efficiently identified and addressed. The error detection and mitigation matrix helps us capture this intersection and identify high-risk regions. For example, using ChatGPT to draft legal motions sits in a high-risk region because errors, such as fake citations, are difficult to detect and even harder to mitigate \cite{Merken2023-pz}. Practitioners must review the entire document, verify citations, and validate arguments, negating any initial efficiency gains. This inadvertently diminishes any efficiency gains that the tool may have provided in the first place. Harm to work practices in the form of increased labor (with no added value) has occurred across several domains where workers had to undertake added labor in the form of repair work \cite{jackson2014rethinking} to address the disruptions caused to decision-making and administrative processes as a result of erroneous outputs \cite{saxena2024algorithmic, saxena2022unpacking, Elish2020-ch, Cheng2022-ya}.

\section{Demonstrating Use through Comparative Cases}\label{sec:results}

In this section, we describe our use of comparative case studies to investigate the matrices' effect on AI concepts. In each of the three examples below, we compare two AI concepts (AI cases) from the same domain to investigate how our matrices can help reveal and clarify risks. We explicitly state the \colorbox{mygreen}{action or task} that the AI concept supports, the \colorbox{mypink}{inference} it makes, and the \colorbox{myblue}{data} that it was trained on. To improve readability, we only discuss matrices that are most relevant to a case study. 

\subsection{Case Study 1: The Allegheny Family Screening Tool (AFST) vs. Hello Baby Predictive Risk Model}

\begin{itemize}
    \item  \textit{\textbf{AFST}}: AI system infers the likelihood of child being \colorbox{mypink}{placed} \colorbox{mypink}{into foster care within two years.} The system uses a cross-section of historical \colorbox{myblue}{public administrative data.} The system output \colorbox{mygreen}{helps a case worker decide} how much attention to give to an individual case.
\end{itemize}
\begin{itemize}
    \item {\textit{\textbf{Hello Baby}}: AI system infers the likelihood of \colorbox{mypink}{placement} \colorbox{mypink}{in foster care by age 3.} The system uses a cross-section of historical \colorbox{myblue}{public administrative data} and proactively offers support by \colorbox{mygreen}{providing information and resources} to families with newborns who may be at risk.}
\end{itemize}
\begin{itemize}
    \item LEGEND: \colorbox{mypink}{inference} \colorbox{mygreen}{action or task} \colorbox{myblue}{data} 
\end{itemize}
\vspace{0.3cm}

The Allegheny County Department of Human Services (ACDHS) in Pennsylvania has implemented two AI systems - 1) the Allegheny Family Screening Tool (AFST) \cite{VaithianathanUnknown-ci} and 2) the Hello Baby predictive risk model (PRM) \cite{VaithianathanUnknown-nl}. AFST is used to assist child welfare hotline workers in deciding whether to investigate calls of child maltreatment. It generates a risk score that indicates the likelihood of the child being placed in foster care within two years. When an allegation of maltreatment is made, a welfare worker collects quantitative information about the family and enters it into AFST to generate a risk score. The caseworker informs their supervisor if they agree or disagree with this score. Risk scores above a certain threshold lead to a mandatory screen-in, and the supervisors must go through a separate override process if they disagree and feel the risk is lower. 

The Hello Baby system introduced a three-tier system of family services to provide early interventions for families in need. It addresses an information gap, that many new parents do not know about publicly available services. The county offers a rich array of services that families with complex needs never access. The program implemented a predictive risk model (i.e., Hello Baby PRM) to identify families with newborns who might benefit the most from resources. It is proactive, pushing an offer of support out to families before problems relating to child maltreatment arise \cite{VaithianathanUnknown-nl}. Participation in the program is voluntary; the families identified as high-risk are not obligated to accept support. At the very least, it makes the families aware of the additional support that is available to them if needed. Below, we assess these two AI concepts and show how the matrices help elicit underlying risks. It also shows that Hello Baby only needs moderate performance to be valuable for welfare workers, which helps avoid several of these risk factors. 

\subsubsection{Required Performance Matrix} 

\begin{figure*}
    \centering
    \includegraphics[width=\linewidth]{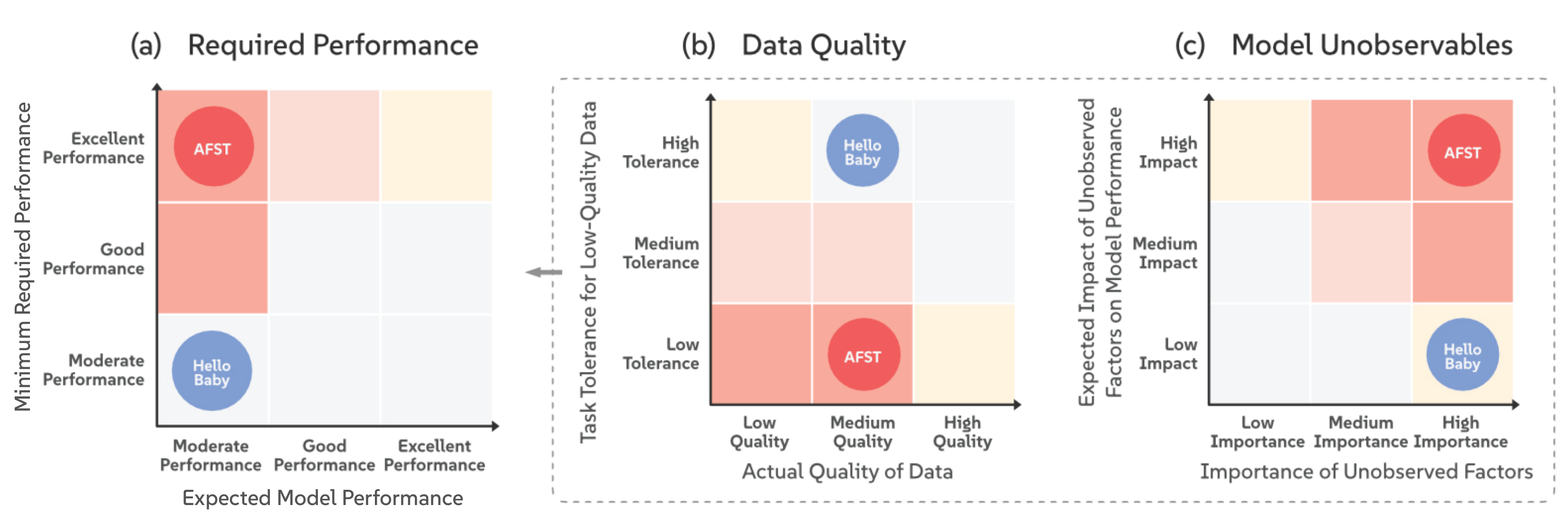}
    \caption{Case Study 1 [AFST vs. Hello Baby] - Data Quality, Model Unobservables, and Expected Performance Analysis}
    \label{fig:afst-hello-1}
    \Description[Case study of AFST and Hello Baby on Data Quality, Model Unobservables, and Expected Performance]{Three-panel matrix comparing data quality, model unobservables, and expected performance between AFST and HelloBaby, with colored circles marking their positions in each matrix.}
\end{figure*}

\begin{figure*}
    \centering
    \includegraphics[width=0.64\linewidth]{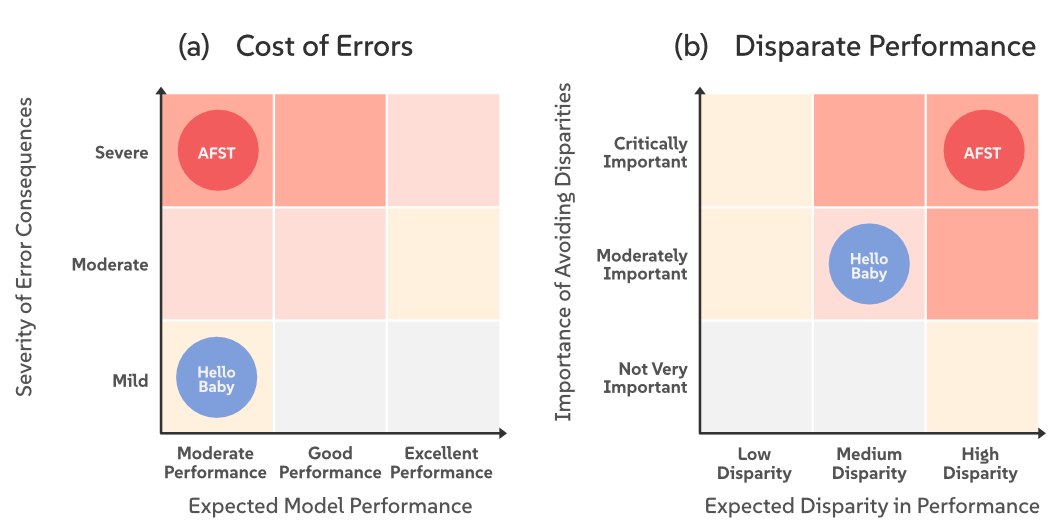}
    \caption{Case Study 1 [AFST vs. Hello Baby] - Cost of Errors and Disparate Performance Analysis}
    \label{fig:afst-hello-2}
    \Description[Case study of AFST and Hello Baby on Cost of Errors and Disparate Performance]{Two-panel matrix comparing Cost of Errors and Disparate Performance between AFST and HelloBaby, with colored circles marking their positions in each matrix.}
    \vspace{0.2cm}
\end{figure*}

{AFST and Hello Baby PRM address different risk levels. AFST is used for high-risk decisions where false positives can separate families unjustly, and false negatives may lead to severe harm, including a child’s death. As shown in Figure~\ref{fig:afst-hello-1}(a), AFST requires excellent model performance to be technically feasible.  In contrast, Hello Baby addresses a lower-risk decision, to communicate or not communicate with new families about available services, and is more tolerant of moderate model performance. False positives only lead to unnecessary information being sent to families, and false negatives don't prevent families from learning about publicly available services. Thus, Hello Baby requires only moderate performance to be valuable, making it technically more feasible.}

{\textbf{Data Quality and Model Unobservables Matrices.}
These matrices facilitate a deeper reflection on expected model performance and may help pinpoint the underlying reasons why lower performance may be expected. Both AFST and Hello Baby rely on public administrative data, which is often sparse, and contains some errors. The data is inconsistently recorded and often out of date. It contains bias from human welfare workers, and this may disadvantage those in poverty or from particular racial or ethnic groups. For example, on December 31, 2014, Pennsylvania state amended the Child Protective Services Law (CPSL) to broaden the category of mandated reporters, leading to inconsistent information in the data used for model training. AFST’s need for excellent performance makes it susceptible to these data quality issues (Figure~\ref{fig:afst-hello-1}(b)). Hello Baby similarly deals with these data quality concerns. However, given its goal — providing early information rather than making high-stakes removal decisions, moderate data quality suffices.}

{Both AFST and Hello Baby deal with unobservable factors (Figure~\ref{fig:afst-hello-1} (c)). AFST predicts the risk of a child entering foster care as a proxy for long-term maltreatment. It looks towards the future. However, welfare workers focus on the present, and on the immediate child safety based on qualitative assessments. This creates a mismatch between the model's predictions and the decision-makers' needs. Hello Baby similarly makes a prediction about the future and the long-term risk of maltreatment. However, this aligns with the decision maker who is focused on a future problem that might be avoided by access to resources and information. Hello Baby, like AFST, faces challenges with unobserved factors, but its universal tier serves as a safety net for families lacking data or experiencing unforeseen events (e.g., immigrant and refugee families). This reduces the impact of lower task performance compared to AFST.}

The matrices reveal that AFST requires excellent model performance to avoid harm, and this level of performance seems unlikely due to the moderate quality data and unobserved factors. This contradiction and mismatch between what is needed and what is realistic could have been identified before developing this system. In contrast, with Hello Baby, the moderate model performance needed and the access to moderate-quality data indicate a much more likely chance of success. Hello Baby’s design is more technically feasible and aligned with its intended goals.

\subsubsection{Cost of Errors Matrix}
{AFST's high error severity and prevalence help explain why this AI system caused harm to vulnerable communities and garnered considerable criticism \cite{Eubanks2018-dk, Ho2022-jj, Ho2023-wy}. Effective AFST implementation requires excellent model performance with minimal consequences for errors, but this level of performance is currently unattainable (Figure~\ref{fig:afst-hello-2} (a)). In contrast, Hello Baby’s errors are less severe. False positives merely result in unnecessary outreach, and false negatives are mitigated by a universal tier of services offered to all families. Even moderate performance thus creates value without causing severe harm. 

\subsubsection{Disparate Performance Matrix}
{AFST also faces high ethical costs due to its disparate performance, as low-income and minority communities are over-represented in public administrative data and could have been foreseen before development (Figure~\ref{fig:afst-hello-2} (b)). Conversely, Hello Baby was designed to fill an information gap rather than dictate punitive measures. Even if it misidentifies some families, they can decline support, and universal services ensure basic resources are accessible to everyone. This approach reduces the ethical cost and the impact of potential disparities. Taken together, Hello Baby creates value even with moderate performance by allowing ACDHS to reach more families and prevent child maltreatment through proactive support.}

In sum, Allegheny County explored a lower-risk, higher-value AI concept after harm to communities and professional practice had occurred and AFST failed to generate value \cite{Kawakami2022-ez, Cheng2022-ya}. Consequently, the county approached the problem of child maltreatment with the intent to lower the risk from the start and find use cases for a predictive model that could still generate value and address the problem of child maltreatment \cite{VaithianathanUnknown-nl}. 

\subsection{Case Study 2: GizmodoBot vs. DuolingoLLM}

    




\vspace{0.3cm}
\begin{itemize}
    \item {\textit{\textbf{Gizmodo Bot}}: AI system \colorbox{mypink}{generates entertainment-focused} \colorbox{mypink}{articles based on a given prompt} and automatically \colorbox{mygreen}{pub-} \\ \colorbox{mygreen}{lishes these articles.} The system uses \colorbox{myblue}{pre-trained LLMs}\\ \colorbox{myblue}{fine-tuned on the company's data.}}
\end{itemize}
\begin{itemize}
    \item {\textit{\textbf{Duolingo LLM}}: AI system \colorbox{mypink}{generates lesson plans and ex-} \\ \colorbox{mypink}{ercises} to \colorbox{mygreen}{inspire course planners} with suggestions. The system uses OpenAI's \colorbox{myblue}{pre-trained LLM,} combined with BirdBrain and Duolingo Max, with \colorbox{myblue}{user learning data.}}
\end{itemize}
\begin{itemize}
    \item LEGEND: \colorbox{mypink}{inference} \colorbox{mygreen}{action or task} \colorbox{myblue}{data}  
\end{itemize}

\vspace{0.5cm}
G/O Media, owner of the science and technology website Gizmodo, developed Gizmodo Bot to automatically write and publish AI-generated articles on entertainment topics \cite{Verma2023-dk}. Gizmodo Bot combines ChatGPT and Google Bard and it’s been fine-tuned with historical Gizmodo articles. The bot writes leisure- or entertainment-centered articles such as, "A Chronological List of Star Wars Movies \& TV Shows." While the system was designed to be used without human oversight, the generated articles were often riddled with factual errors \cite{Verma2023-dk}. Employees complained that the AI-generated content diminished their journalistic integrity and credibility.

\vspace{0.2cm}
In contrast, the language learning app Duolingo developed a backend tool to assist its education experts with developing lesson plans \cite{Henry2023-vm}. Duolingo partnered with OpenAI, using their LLMs for several features such as BirdBrain \cite{Bicknell2020-it} and Duolingo Max \cite{Team-Duolingo2023-bs}. In analyzing this case, we specifically focus on the use of LLMs internally to support professionals. It is important to note that the AI integration did result in the company laying off 10\% of its contractors \cite{De-Vynck2024-ua}. Below, we analyze these two cases as if they were AI concepts. We use the matrices and highlight why one led to harm, while the other was able to augment workers’ practices using moderate AI performance.

\subsubsection{Required Performance Matrix}

\begin{figure*}
    \centering
    \includegraphics[width=\linewidth]{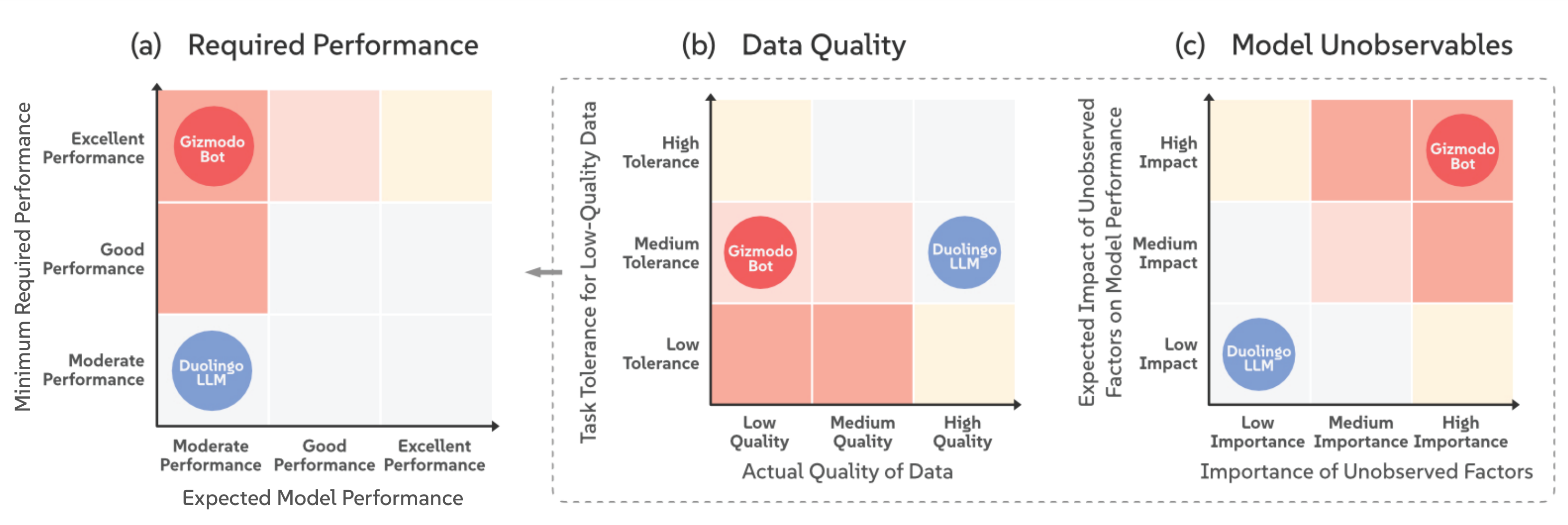}
    \caption{Case Study 2 [Gizmodo Bot vs. Duolingo LLM] - Data Quality, Model Unobservables, and Expected Performance Analysis}
    \label{fig:gizmodo-duolingo-1}
    \Description[Case study of Gizmodo Bot and Duolingo LLM on Data Quality, Model Unobservables, and Expected Performance]{Three-panel matrix comparing Data Quality, Model Unobservables, and Expected Performance between Gizmodo Bot and Duolingo LLM, with colored circles marking their positions in each matrix.}
\end{figure*}

\begin{figure*}
    \centering
    \includegraphics[width=0.64\linewidth]{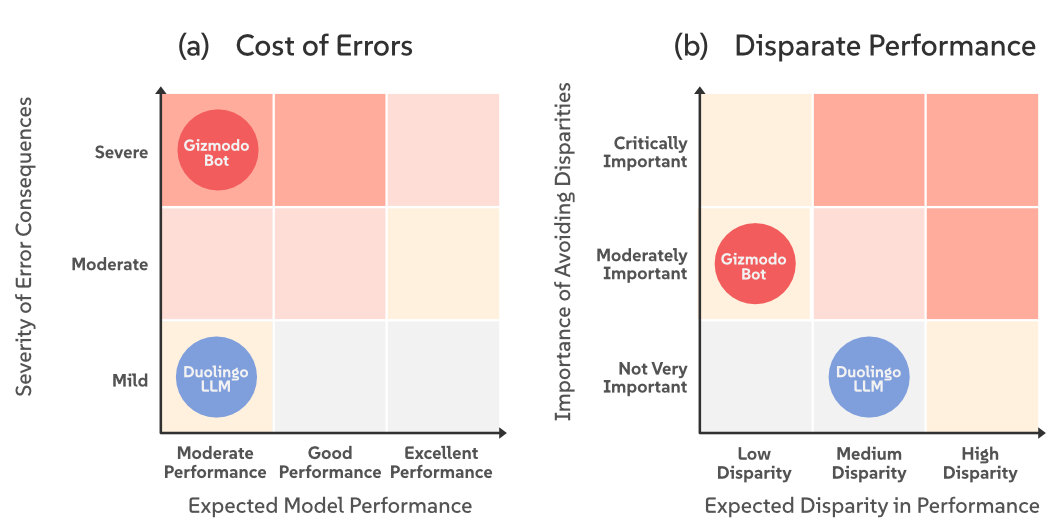}
    \caption{Case Study 2 [Gizmodo Bot vs. Duolingo LLM] - Cost of Errors and Disparate Performance Analysis}
    \label{fig:gizmodo-duolingo-2}
    \Description[A case study of Gizmodo Bot and Duolingo LLM on Cost of Errors and Disparate Performance Analysis]{Two-panel matrix comparing Cost of Errors and Disparate Performance between Gizmodo Bot and Duolingo LLM, with colored circles marking their positions in each matrix.}
    \vspace{0.3cm}
\end{figure*}

{Gizmodo Bot’s use case demanded accurate, publish-ready content. This high standard implied that it needed excellent model performance to be technically feasible. However, Gizmodo Bot struggled to understand the context and address the temporality of new information \cite{Xu2024-jq, Wallat2024-nw}. For example, various online articles focused on Star Wars struggled with the differences between release order, chronological order, and naming order \cite{Ujlaki2024-ok}. Addressing this appropriately requires a common sense understanding of time that is beyond LLM capabilities. In contrast, Duolingo utilizes LLMs to quickly generate large amounts of content, which teaching experts then refine into lesson material. Experts anticipate errors and often select the best three sentences out of ten generated by the LLM. Given that quality control happens at the human level where the experts are able to quickly filter out irrelevant or low-quality results, moderate AI performance is acceptable and still valuable. Duolingo’s approach tolerates less-than-perfect outputs where human experts transform these outputs into finished products.}

{\textbf{Data Quality and Model Unobservables Matrices.}
These matrices facilitate a deeper reflection on expected model performance and may help pinpoint the underlying reasons why lower performance may be expected. For Gizmodo Bot, data quality and missing context matter greatly. Cultural knowledge, temporal relevance, and factual accuracy are unobservable factors that LLMs struggle with when left unchecked. Here, journalists and editors play a crucial role in fact-checking, situating the draft within current affairs, and adding expert quotes to contextual critical parts. The G/M media system was designed to publish AI-generated articles without human oversight. With no human oversight, the system’s moderate-quality data and inability to capture real-world nuances meant it could never achieve the excellent performance needed.}

{For Duolingo LLM, given that experts are able to quickly filter out irrelevant or low-quality results, the model has the potential to be useful even with moderate-quality training data (Figure~\ref{fig:gizmodo-duolingo-1} (b)). However, there are critical unobservable factors associated with developing lessons -- transforming LLM-generated content into learning exercises requires pedagogical knowledge (e.g., following a learning trajectory, setting appropriate difficulty levels, assessing cultural or linguistic biases, etc.). Duolingo's approach accounts for these model unobservables and further demonstrates how moderate AI performance can still create value.}


\subsubsection{Cost of Errors Matrix} 
{Gizmodo Bot was designed to write entertainment-focused content and had a low severity of errors. However, a moderate amount of errors should have been expected because LLMs are known to hallucinate. With no mitigation strategies in place, this consequently increased the severity and caused reputational harm to the organization and harm to workers’ practices. Given that similar products had already failed at media and news companies, there were lessons to be learned \cite{Sato2023-bm, Christian2023-ap}. The editor, as quoted by the Washington Post, shared they have never had to deal with this basic level of incompetence \cite{Verma2023-dk}. In contrast, Duolingo’s approach inherently accepts error-prone raw material. Because experts intervene before any content reaches learners, the severity of errors is low. Mistakes are filtered out during the curation process, meaning the cost of errors remains manageable. This allows the AI to still contribute value despite imperfect outputs. 

\subsubsection{Cost of Disparate Performance Matrix} 
{Gizmodo Bot was designed to write mainstream media content as a way to attract search traffic and generate ad revenue where the expected disparateness in performance is minimal. If it were repurposed to also generate content on more niche topics with sparse data sources (e.g., independent films), worse performance would be expected due to a higher likelihood of hallucinations. Similarly, for Duolingo’s use case, disparateness in performance across languages should be expected given unequal training data. However, there is a low cost of disparate performance because teaching experts expect errors and treat AI output as raw material that they must mold and refine.}

In sum, Gizmodo Bot was designed to assist journalists with low-stakes tasks, freeing them to focus on more important stories. However, it caused harm due to a lack of attention to workers' practices. In contrast, Duolingo's approach accepts that AI will make mistakes (i.e., moderate AI performance) but still effectively supports expert labor. It benefited from setting realistic expectations and integrating human expertise into the workflow. That is, responsible AI practitioners should explore innovations that augment workers' practices with moderate AI performance rather than seeking to fully automate an expert task that would inadvertently need excellent performance.



\subsection{Case Study 3: Thomson Reuters’ \textit{CLEAR} for Retail and Financial Services vs. \textit{Fraud Detect} for Social Services}

\begin{itemize}
    \item \textit{\textbf{Fraud Detect for Social Services}}: \colorbox{mygreen}{Approve, deny, or qua-}\\ \colorbox{mygreen}{rantine citizens' applications} for public benefits based on the \colorbox{mypink}{likelihood of fraudulent behavior} assessed using \colorbox{myblue}{pu-} \colorbox{myblue}{blic administrative data, citizens’ social media data, credit} \\ \colorbox{myblue}{reports, and housing records.}
\end{itemize}
\begin{itemize}
    \item \textit{\textbf{CLEAR for Retail and Financial Services}}:{\colorbox{mygreen}{Block or qua-} \\ \colorbox{mygreen}{rantine financial transactions} based on the \colorbox{mypink}{likelihood of} \\ \colorbox{mypink}{fraudulent behavior} assessed using the customers’ \colorbox{myblue}{finan-} \colorbox{myblue}{cial transactions data.}}
\end{itemize}
\begin{itemize}
    \item LEGEND: \colorbox{mypink}{inference} \colorbox{mygreen}{action or task} \colorbox{myblue}{data} 
\end{itemize}

Thomson Reuters is an information conglomerate with online and print business, tax and accounting services, and law and compliance services. The company has also developed a suite of fraud detection AI tools to support its private and public sector clients. Thomson Reuters first developed these tools for their retail and financial sector customers to help them identify fraudulent transactions and conduct identity verification. Witnessing success in these domains, the company has developed an entire corporate arm on "Risk \& Fraud" \cite{ReutersUnknown-lu} and acquired FraudCaster in 2020 (renamed to Fraud Detect) extending fraud detection AI services to public agencies such as unemployment, child welfare services, Medicaid programs, and policing. CLEAR for retail and financial services analyzes the clients’ financial transaction data and verifies customer identities to detect anomalies. The tool has access to high-quality financial data and has a higher tolerance for low-quality data where it may freeze transactions for a professional to review and approve such that even moderate AI performance generates value for the clients. On the other hand, Fraud Detect uses less reliable data sources (public administrative data, social media data, credit reports, and housing records) to predict if public benefits applications (i.e., unemployment, SNAP, Medicare, and Medicaid benefits) are fraudulent and automatically denies or suspends them if deemed fraudulent. We analyze these two cases as AI concepts and through the matrices show why Fraud Detect faced multiple risks from the start while CLEAR was able to avoid them. We also note that CLEAR has been criticized for unethical data collection and reappropriated to be used by law enforcement. Our focus here is specifically on the use of CLEAR in financial services for detecting financial fraud. 

\begin{figure*}
    \centering
    \includegraphics[width=\linewidth]{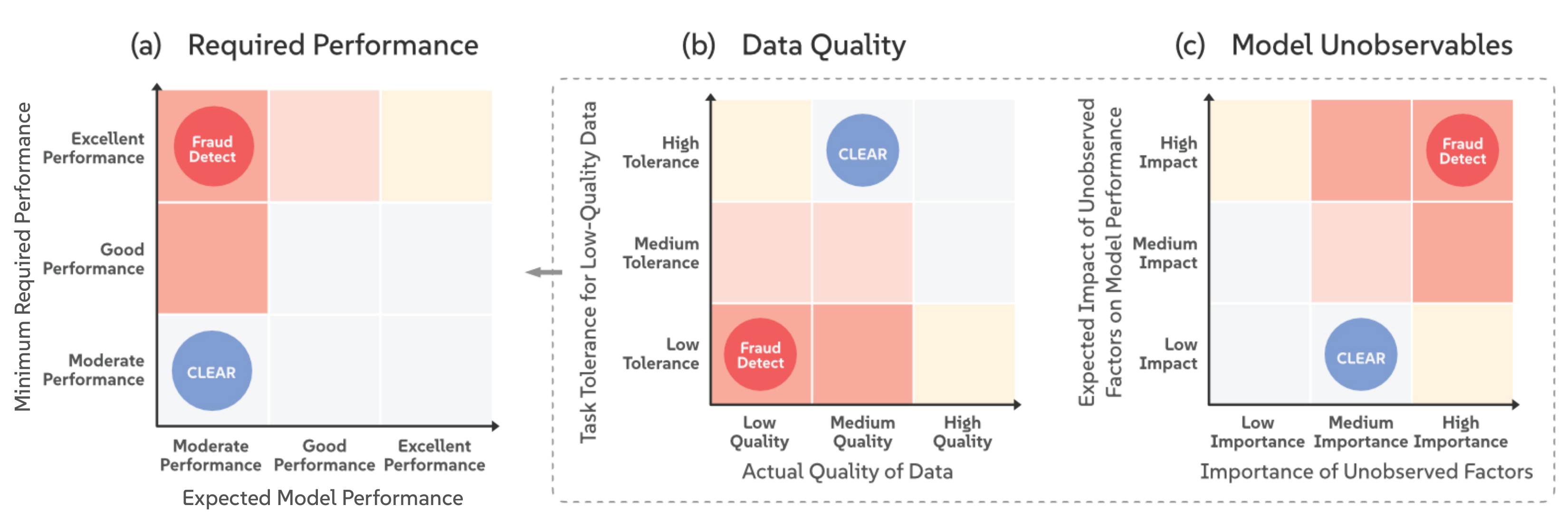}
    \caption{Case Study 3 [CLEAR vs. Fraud Detect] - Data Quality, Model Unobservables, and Expected Performance Analysis}
    \label{fig:clear-fraud-1}
    \Description[Case study of Thomson Reuters' CLEAR and Fraud Detect service on Expectation of Errors and Error Detection \& Mitigation]{Three-panel matrix comparing Data Quality, Model Unobservables, and Expected Performance between CLEAR and Fraud Detect service, with colored circles marking their positions in each matrix.}
\end{figure*}

\begin{figure*}
    \centering
    \includegraphics[width=0.64\linewidth]{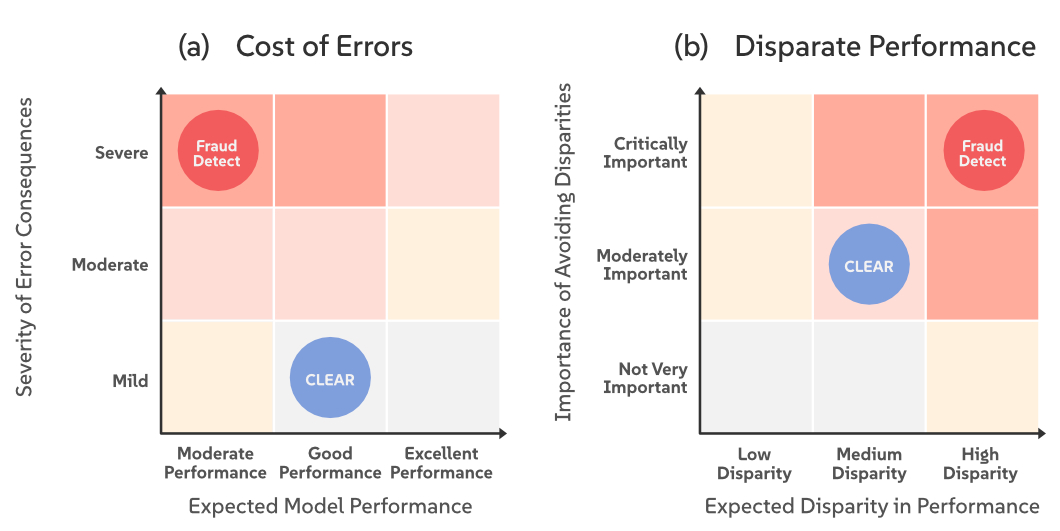}
    \caption{Case Study 3 [CLEAR vs. Fraud Detect] - Cost of Errors and Disparate Performance Analysis}
    \label{fig:clear-fraud-2}
    \Description[A case study of Thomson Reuters' CLEAR and Fraud Detect service on Cost of Errors and Disparate Performance]{Two-panel matrix comparing Cost of Errors and Disparate Performance between CLEAR and Fraud Detect service, with colored circles marking their positions in each matrix.}
\end{figure*}

\subsubsection{Required Performance Matrix}
{CLEAR is well-suited to achieve moderate expected performance in detecting fraudulent transactions, similar to an email spam filter that flags suspicious activity for further review. This AI capability has been successful across the financial sector, blocking transactions from known fraudulent entities and flagging unusual transactions. By contrast, Fraud Detect’s expected performance needs to be excellent due to the high-risk domain (i.e., public benefits), yet the system may struggle to achieve the required model performance given the constraints. Figure~\ref{fig:clear-fraud-1}(a) illustrates the mismatch between what is needed and what is realistic. Fraud Detect’s inability to meet these expectations has led to denying more than half of legitimate claims in California, leaving about 600,000 vulnerable people without essential services \cite{Quinlan2024-iy, epic_report_2023}. Below, we further reflect upon expected model performance using the Data Quality and Model Unobservables matrices to better understand the underlying constraints that impact performance.}

\vspace{0.2cm}
{\textbf{Data Quality and Model Unobservables Matrices.} The data quality for CLEAR’s financial fraud detection is high, which supports moderate expected performance. The system can rely on relatively clean, structured transaction data, making anomaly detection straightforward and allowing modest performance to still deliver value. Comparatively, Fraud Detect relies on public administrative data that has serious limitations, along with biased social media, credit reporting, and housing data \cite{Allard2018-hu, Logan2016-of, Botsis2010-ac}. For example, information about clients using public services can be inconsistent and contradictory because agencies maintain separate records that better represent their own operations, rather than clients’ overall circumstances \cite{Holten-Moller2020-fe, Moon2024-si, Saxena2020-qh}. Given the high-risk task, this Fraud Detect has a low tolerance for low-quality data. Yet high-quality data is needed to create value; a critical mismatch for the task, as can be observed in Figure~\ref{fig:clear-fraud-1} (b).}

\vspace{0.3cm}
{Even with CLEAR, given the emerging patterns of fraud, some model unobservables are always present (e.g., features emerging from novel fraud schemes, coordinated fraud, etc.). Therefore, moderate AI performance should be expected which is enough to provide value to stakeholders and prevent harm. For Fraud Detect, civil servants consider unobserved factors not recorded in the data, such as changes in life circumstances, efforts to gain employment, and number of dependents, among other factors. Ignoring them can lead to significant harm, as evidenced by a similar AI system in the Netherlands that caused substantial harm and led to the cabinet’s resignation \cite{Rao2022-ad} (Figure~\ref{fig:clear-fraud-1}(c)).  The limitations of Fraud Detect prevent achieving the necessary performance and reveal a critical mismatch between its intended and pragmatic use (Figure~\ref{fig:clear-fraud-1} (a)). Given these prior failures, harm could have been anticipated. Fraud Detect has denied more than half of legitimate claims leaving about 600,000 vulnerable people without essential services \cite{Quinlan2024-iy, epic_report_2023}.}

\subsubsection{Cost of Errors Matrix}
{In financial services, AI can detect fraudulent transactions significantly faster than a human professional. In addition, given that it follows a block and quarantine approach similar to email spam, the severity of errors is lower where quarantined transactions need approval (Figure~\ref{fig:clear-fraud-2} (a)). Here, even moderate performance creates significant value for the stakeholders. Comparatively, Fraud Detect has a higher severity of errors (vulnerable people are being denied services) and a higher prevalence of errors due to data limitations and model unobservables. It was not technically feasible to begin with, and a high number of costly errors should have been expected. 

\subsubsection{Disparate Performance Matrix}
{Moreover, CLEAR's financial fraud detection aims to identify anomalous customer behavior and transactions. Since many companies develop strategies targeting specific demographic groups, class imbalance in the data is expected, leading to moderate performance disparities (Figure~\ref{fig:clear-fraud-2} (b)). On the other hand, detecting fraud in applications for public benefits has a significantly higher likelihood of disparate performance due to biases in the training data: different demographic groups utilize public benefits programs at very different rates \cite{Wang2024-cj, Chouldechova2017-aj, Kleinberg2016-fd}. There is also a much higher cost of disparate performance since some demographic groups, especially low-income and minority communities, may be denied essential services.}

In sum, we highlight how an AI capability that is ported across domains can itself introduce new risks that can go undetected if not examined during the early stages. Moreover, moderate AI performance in detecting fraud may have sufficed to create value for retail and financial services but excellent model performance is needed in social services because the risk of causing harm to vulnerable people is very high.

\section{Discussion}

\subsection{Scaffolding Exploration of AI’s Design Space through a Risk Framework}
Recent work has argued for the need to broaden innovation teams' explorations of AI’s design space~\cite{kawakami2024situate, Stapleton2022-eb, Wang2024-cj, yildirim2023creating}. For example, work on Responsible AI has urged teams to explore opportunities for AI use beyond high-risk, consequential decisions~\cite{kawakami2024situate, Raji2022-ls, Stapleton2022-eb, Wang2024-cj}. HCI researchers have developed toolkits to support broader ideation and help innovation teams find the right things to design (e.g.,~\cite{yang2019sketching,yildirim2023creating}). They also recognized that due to the inherent uncertainty associated with AI inference, AI is a uniquely difficult design material to work with~\cite{yang2020re}. Better resources and processes are required for scaffolding the risk assessment process for AI concepts, before selecting which AI concept to move forward to prototyping. Our risk matrices support two critical goals: 1) facilitate communication about project risks across team members (e.g., UX practitioners and data scientists) who may bring complementary knowledge and expertise relevant to risk assessment, and 2) visualize high-risk regions, assess relative riskiness of AI concepts, and help refine them to reduce risk while retaining value.

\subsubsection{Supporting communication and collaboration to identify AI Mismatches} 
Research in HCI and adjacent communities such as FAccT and AIES has recognized the need for the AI innovation process to be collaborative, interdisciplinary, and cross-functional --- engaging diverse expertise in the assessment of potential risks and benefits (e.g.,~\cite{delgado2023participatory, deng2023investigating,kawakami2024situate,kuo2023understanding,subramonyam2021towards}). However, to support such collaborative processes, conversational resources are needed that make technical concepts more accessible and facilitate deliberation and sharing of information. We adopted an approach that captures technical terms in a descriptive form where the riskiness of AI concepts can be depicted as a function of two orthogonal risk dimensions that together reveal AI Mismatches. Our approach offers an intuitive way to analyze critical factors and identify dimensions along which the AI concept can be refined to reduce risk. The matrices also ask data scientists to adopt a more human-centric approach to model evaluation and focus on how well the AI needs to perform to be useful rather than just predictive accuracy on a proxy metric~\cite{bansal2019beyond,guerdan2023ground,hutchinson2022evaluation, Raji2022-ls, Wang2024-cj, yildirim2023creating}. Similarly, they can think about errors in terms of severity, prevalence, and humans' ability to detect them instead of just precision and recall.

We envision that tools like our matrices may also support stakeholder alignment on project goals and acceptable risk tolerance. Ongoing communication during early concept analysis, redesign, and pre-deployment is essential, as AI projects often shift or change course when high-level objectives are iteratively translated into computational problems~\cite{Kross2021-fb,passi2019problem}. Here, communication tools like these matrices can help convey to diverse stakeholders how much an AI project has drifted along risk dimensions since the early concept analysis.

\subsubsection{Matrices as a visualization and reasoning tool to assess tradeoffs}
The matrices serve as a canvas to visualize and assess the relative riskiness of different AI concepts. These matrices can be used to make informed tradeoffs. For instance, consider the tradeoff between model performance and complexity---a team might focus on improving model performance (to lower error prevalence) by using more complex modeling approaches. However, this might make it harder to detect and mitigate errors, moving the AI concept into a high-risk region on that matrix. Overall, the matrices provide a structured approach to assessing and negotiating early-stage AI project risks. Stakeholders can see, visually, where risk lies and assess whether developing high-risk AI projects would generate value, given associated risks.

\subsection{Implications for Responsible AI}
Our study provides some important implications for responsible AI practice. First, we reflect on some early insights that suggest that a focus on \textit{moderate AI performance} may help innovation teams avoid risk factors that can result in critical functionality issues down the line. Next, we discuss some implications for the worker-centered design of AI systems. Finally, we draw caution against uncritically transferring or adapting AI concepts across domains.   

\subsubsection{Moderate AI Performance may support responsible AI innovations}
In alignment with prior work~\cite{yildirim2023creating}, our analysis of widely adopted AI applications, along with case studies, shows that moderate AI performance can often create value for stakeholders while minimizing harm. These applications were able to avoid several risk factors across the matrices. Focusing on moderate AI performance implies three key expectations: 1) higher error prevalence means the consequences of errors can't be severe, as too many costly errors would lead to critical failures. This might push an innovation team to explore lower-risk use cases, 2) higher error prevalence requires users to have a greater tolerance and expectation for AI errors for the tool to be valuable, and finally, 3) since the users expect errors and the severity of errors needs to be low, the cost to mitigate these errors is also lower. We believe that this may be a promising and pragmatic approach to responsible AI and requires deeper exploration. First, this may help innovation teams identify lower-risk, higher-value opportunities. Once a prototype shows value, they can iteratively improve its performance, if necessary. Second, it may also help establish the baseline for appropriate trust and reliance on the AI tool because the expectation is that it will make mistakes. And finally, it may also help avoid downstream costs related to error or harm detection and mitigation; a critical factor that frequently leads to AI project failure because the costs outweigh value creation.

\subsubsection{A worker-centered approach to AI innovation}
Several high-expertise tasks are characterized by uncertainties that require expert discretion. AI systems designed to automate such tasks must perform exceptionally well to create value. Gizmodo Bot was intended to support journalists by handling lower-stakes tasks and allowing them to focus on more significant stories. However, it ended up causing harm due to a poor consideration of workers’ practices. CNET and Men’s Journal experienced similar issues, needing to rewrite error-ridden AI-generated articles \cite{Sato2023-bm, Christian2023-ap}. Comparatively, similar to the DuoLingo use case, we are seeing AI concepts emerge across domains such as public services \cite{Elisco2024-bu} and healthcare \cite{Shanks2024-bt} that operate with moderate performance and still support expert labor effectively. Recognizing that AI will make mistakes and that human experts are needed to address uncertainties and refine AI outputs allows for more effective and creative integration of AI into workflows that augment workers’ practices. This further helps build appropriate trust and reliance in the AI system because the workers have a clear understanding of the failure modes and when to expect them in practice.

\subsubsection{Analogical inspiration for AI innovation can lead to harm}
Analogical thinking, which involves drawing parallels between different domains to inspire innovation~\cite{hope2017accelerating,kittur2019scaling,kang2022augmenting}, has helped spur AI innovation. By transferring solutions from one context to another, innovators can tackle new problems by leveraging knowledge and concepts from unrelated fields. For instance, early AI adoption in the public sector was inspired by healthcare, where statistical predictions outperformed clinical judgment on various tasks~\cite{AEgisdottir2006-op,Dawes1989-um,Meehl1954-ee}. However, we draw caution against uncritically porting AI capabilities across domains, as different contexts may introduce new risks that can go undetected. In the third case study, the matrices helped systematically examine two seemingly similar AI concepts and uncover risk factors that placed FraudDetect in higher-risk regions relative to CLEAR.

\subsection{Limitations} 
Our focus was on identifying inherent risk factors in AI concepts at the earliest stages of AI innovation. Consequently, our analysis captures high-risk regions of the problem-solution space for AI and the harms that have occurred in those regions. The matrices presented in this paper focus on gaps between task performance requirements and realistic model performance and are not meant to capture downstream harms resulting from technology misuse (e.g., technology-facilitated intimate partner violence \cite{bellini2023paying}) or unintended privacy violations (e.g., recommender systems revealing users’ personal information \cite{tufekci2015algorithmic}). A more thorough concept assessment is needed that balances risks with benefits once the innovation team has rejected high-risk/low-value ideas. Finally, our goal with this study was to develop a descriptive (and not prescriptive) approach to early-stage risk assessment that allows researchers and innovation teams to deliberate on risk factors and be able to visualize and communicate them with other stakeholders. A prescriptive approach that attempts to precisely define risk within every quadrant risks limiting interpretive flexibility for teams to assess project risks in their specific contexts.

\section{Conclusion and Future Directions}
This paper offers an initial step towards understanding the factors that influence AI Mismatches --- discrepancies that indicate an AI concept might be infeasible to implement in a way that both creates value and minimizes harm. Our approach helps teams consider several perspectives on unintended harms for a given concept and increase the ability to foresee potential harms and assess ways to minimize them before system development. It can support teams in steering their concepts toward ‘safer’ zones, where a more balanced approach can be taken between what a concept aims to achieve and what is realistically possible. We present a set of comparative case studies where we use our approach to uncover key AI Mismatch factors that influence the gap between AI capabilities and task requirements. We also identify areas for future study related to AI Mismatch, particularly around moderate model performance. Future work can help to discover additional design resources that can support an innovation team in ensuring that AI projects do not simply avoid harm, but create value for stakeholders. 


\begin{acks}
This research is supported by the National Science Foundation under Grant No. (2007501), the Presidential Postdoctoral Fellowship Program (PPFP) at Carnegie Mellon University, the Digital Transformation and Innovation Center at Carnegie Mellon University sponsored by PwC, and the UL Research Institutes through the Center for Advancing Safety of Machine Intelligence. Any opinions, findings, conclusions, or recommendations expressed in this material are those of the authors and do not necessarily reflect the views of our sponsors or research partners. We thank our anonymous reviewers whose valuable feedback and suggestions helped improve this manuscript.
\end{acks}

\bibliographystyle{ACM-Reference-Format}
\bibliography{refs, paperpile}

\appendix

\newpage

\section{Step-by-Step Workflow for Applying the AI Mismatch Matrices}~\label{sec:workflow}
This framework is designed to guide AI innovation teams through a structured, reflective process where the model's performance is judged by its ability to support human needs. It adopts a human-centric definition of model performance \cite{parasuraman2000model, yang2019unremarkable, lai2021towards}. We define model performance holistically in terms of the \textit{model's ability to perform a task that fulfills a human need}, instead of defining it based on ML metrics (e.g., predictive accuracy) that capture predictive performance on historical data but not necessarily the model's utility in supporting a real-world human task.

\subsection{Define the AI Concept and Associated Task}
Begin by specifying the human need your AI concept is designed to fulfill. This step focuses on articulating the problem in terms of real-world impact and value, ensuring that the objective goes beyond traditional metrics to emphasize utility and human benefit.

\noindent \textbf{Identify the Problem} -- Clearly specify the problem the AI concept aims to solve by stating the \textit{action or task} that the AI concept supports, the \textit{inference} it makes, and the \textit{data} that it is trained on.

\noindent \textbf{Set Goals for Value Creation} -- Determine the real-world value you expect the AI concept to deliver (e.g., increased efficiency, improved safety, better decision-making).


\subsection{Position Design Concepts within the Three Core Matrices}
Turn your attention to the core matrices that examine the AI concept's overall ability to deliver value. As needed, use the supporting matrices to structure your thinking about where to place a design concept within the core matrices.

\vspace{0.3cm}
\noindent\textbf{Required Performance Matrix}
\begin{itemize}
    \item QUESTION: ``Can the expected model performance sufficiently deliver the intended value?”
    \item ACTION: Compare the expected performance—framed in terms of its real-world impact—with the minimum threshold for performance needed to effectively serve the human need.
\end{itemize}

\noindent\textbf{Disparate Performance Matrix}
\begin{itemize}
    \item QUESTION: “Can we ensure equitable performance across different user groups?”
    \item ACTION: Assess whether equitable performance is essential for a given AI use case. If so, assess how the model performs across various demographic groups to assess whether it consistently meets stakeholder needs.
\end{itemize}

\noindent\textbf{Cost of Errors Matrix}
\begin{itemize}
    \item QUESTION: “Is the expected model performance sufficient to minimize harm when errors occur?”
    \item ACTION: Evaluate the potential consequences of errors, considering their impact on humans, to determine if the expected performance is sufficient for minimizing harm when the AI concept makes mistakes.
\end{itemize}


\subsection{Supplementary Analysis with Supporting Matrices}
If uncertainties or potential mismatches emerge during your core assessments, consult the supporting matrices to gather more detailed insights. These helper tools provide additional context on factors that might affect (human-centric) model performance.

\vspace{0.2cm}
\noindent\textbf{Data Quality Matrix}
\begin{itemize}
    \item Assess the relevance, completeness, and fairness of your data to understand if any limitations might skew model performance and the ability to support the human task.
\end{itemize}

\vspace{0.2cm}
\noindent\textbf{Model Unobservables Matrix}
\begin{itemize}
    \item Determine if any critical features or contextual information is absent that could lead to performance issues and undermine the model’s utility in practice.
\end{itemize}

\vspace{0.2cm}
\noindent\textbf{Expectation of Errors \& Error Detection/Mitigation Matrices}
\begin{itemize}
    \item Examine user perceptions and impact of errors, then evaluate the effort required for identifying and fixing errors—both of which determine the operational severity of the errors.
\end{itemize}

\vspace{0.3cm}
\subsection{Integrate Findings, Deliberate Over the AI Concept's Feasibility}
Looking across the core matrices, come together as a team to strategize about potential refinements to current design concepts, or ideas for new concepts, to reduce risk before further development. By embedding human impact at every step, teams can ensure that their AI innovations are both technically robust and practically valuable in real-world contexts.

\vspace{0.5cm}
\section{AI Mismatch Matrices: Reference Guide}~\label{sec:flyer}

The following page presents the "AI Mismatch Matrices: Reference Guide," a succinct, one-page document designed to assist teams in identifying and addressing areas of concern or uncertainty related to AI implementations. This guide facilitates focused discussions by highlighting key issues and suggesting potential solutions. Teams can download and print this guide from the Supplemental Material section for easy reference.

\begin{figure*}
    \includegraphics[height=\textheight]{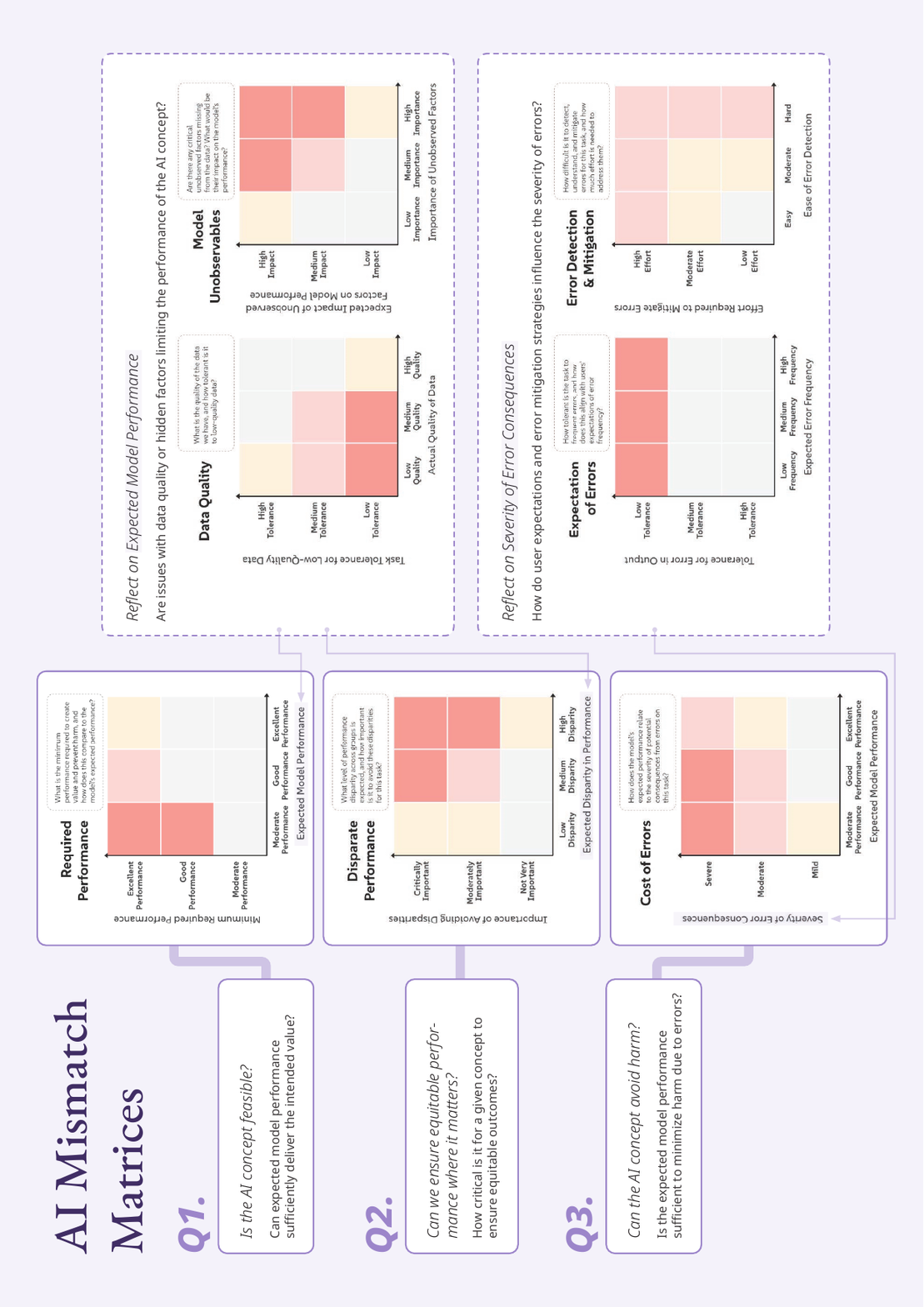}
    \Description[AI Mismatch flyer image]{This figure depicts an overview of the relationship between AI Mismatch matrices.}
\end{figure*}


\end{document}